\documentclass[11pt]{iopart}

\usepackage{iopams}
\usepackage{cases}
\usepackage{makeidx}
\makeindex
\usepackage {graphicx}
\usepackage {subfigure}
\usepackage{color}

\newcommand{\braket}[2]{\left\langle #1 \vphantom{#2} \right|
\left. #2 \vphantom{#1} \right\rangle} 

\begin{document}

\title[Detection at the same place of two electrons.]{Time-dependent exchange and tunneling: detection at the same place of two electrons emitted simultaneously from different sources}

\author{\bf{D Marian}$^{1}$, \bf{E Colom\'{e}s}$^{2}$, and \bf{X Oriols}$^{2}$}
\address{$^1$Dipartimento di Fisica dell'Universit\`a di Genova and INFN sezione di Genova, Via Dodecaneso 33, 16146 Genova, Italy}
\address{$^2$Departament d\rq{}Enginyeria Electr\`{o}nica, Universitat Aut\`{o}noma de Barcelona, 08193, Bellaterra, SPAIN}

\begin{abstract}
Two-particle scattering probabilities in tunneling scenarios with exchange interaction are analyzed with quasi-particle wave packets. Two initial one-particle wave packets (with opposite central momentums) are spatially localized at each side of a barrier. After impinging upon a tunneling barrier, each wave packet splits into transmitted and reflected components. When the initial two-particle anti-symmetrical state is defined as a Slater determinant any type of (normalizable) one-particle wave packet, it is shown that the probability of detecting two (identically injected) electrons at the same side of the barrier is different from zero in very common (single or double barrier) scenarios. In some particular scenarios, the transmitted and reflected components become orthogonal and the mentioned probabilities reproduce those values associated to distinguishable particles. These unexpected non-zero probabilities are still present when non-separable Coulomb interaction or non-symmetrical potentials are considered. On the other hand, for initial wave packets close to Hamiltonian eigenstates, the usual zero two-particle probability for electrons at the same side of the barrier found in the literature is recovered. The generalization to many-particle scattering probabilities with quasi-particle wave packets for low and high phase-space density are also analyzed. The far-reaching consequences of these non-zero probabilities in the accurate evaluation of quantum noise in mesoscopic systems are briefly indicated. 
\end{abstract}


\maketitle


\section{Introduction}
\label{sec1}

The ultimate reason why the quantum theory gives rise to a host of puzzling and fascinating phenomena is because many-particle quantum systems are defined in a high-dimensional and abstract configuration space. For example, in a system of identical particles, only those wave functions whose probability density in the configuration space remains unchanged under permutations of particles are acceptable. When this happens, it is said that the system has exchange interaction. One consequence of the exchange interaction is the Pauli repulsion, that forces electrons to avoid common positions.\\

\begin{figure}[h]
\includegraphics[scale=0.43]{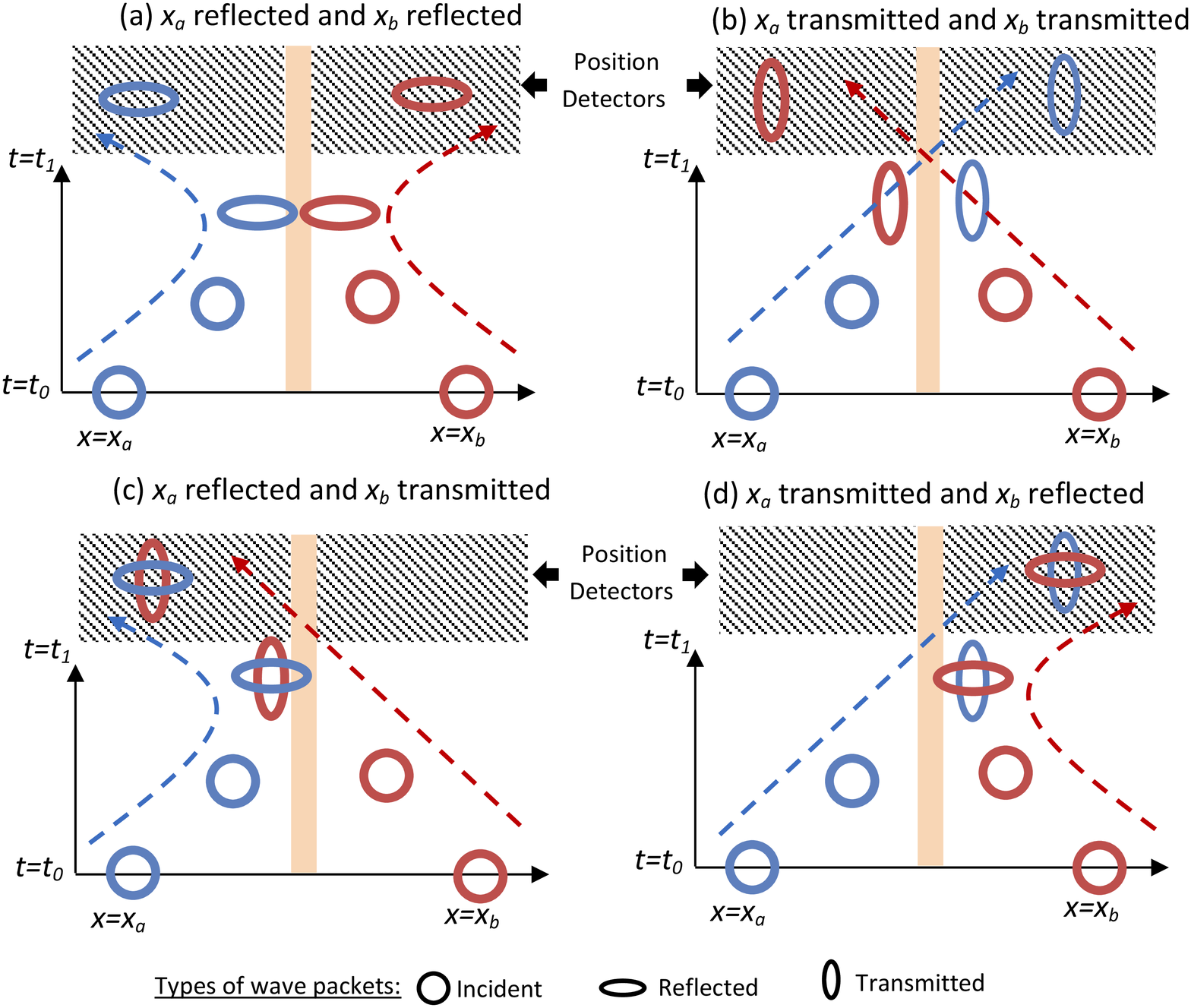}
\caption{Two identically injected wave packets from the left $x_a$ and from the right $x_b$ of a scattering barrier. Solid regions represent the barrier region and shaded regions represent the particle detectors. (a) and (b) each particle is detected on a different side of the barrier  at final time $t_1$ when the interaction with the barrier has almost finished. (c) and (d) both particles are detected on the same side of the barrier. }
\label{fig1}       
\end{figure}

 A typical scenario for discussing exchange and tunneling is shown in the scheme of \fref{fig1}. Two electrons with the same energy and opposite momentum are injected simultaneously from two different sources. During the interaction with a tunneling barrier, the initial one-particle state in the physical space splits into a transmitted and a reflected part. At the final time, transmitted and reflected components coincide in the same spatial region and the Pauli repulsion becomes relevant. Within the (Landau) Fermi liquid theory \cite{Landau1,Landau2,Landau3}, the type of scattering experiments depicted in \fref{fig1} are analyzed by assuming that the (quasi-particle) electrons are described by a one-particle mono-energetic scattering state \cite{Buttiker2,Buttiker3}. The creation and annihilation operators in the second quantization formalism provide a very elegant and powerful formalism to construct the Slater determinant combination. The (anti-symmetrical) initial many-electron state with one electron at each side of the barrier is defined by $ | \Psi \rangle= \hat{a}^{\dagger}_{L} \hat{a}^{\dagger}_{R} | 0 \rangle$. The scattering theory for mono-energetic states predicts that  the probability of finding one electron on the left and one electron on the right of the barrier is:
 
\begin{eqnarray}
\mathcal{P}^{S}_\mathcal{LR} =  | \langle 0 | \hat{b}_{L} \hat{b}_{R} \hat{a}^{\dagger}_{L} \hat{a}^{\dagger}_{R} | 0 \rangle |^{2} =1, \label{p_lr}
\label{B_LR}
\end{eqnarray}
where the upperindex $S$ indicates scattering formalism. Equivalently, the probability of finding both electrons on the left is: 
\begin{eqnarray}
\mathcal{P}^{S}_\mathcal{LL} =  |\langle 0 |  \hat{b}_{L}  \hat{b}_{L}  \hat{a}^{\dagger}_{L} \hat{a}^{\dagger}_{R} | 0 \rangle |^{2} = 0 \label{p_ll}
\label{B_LL}
\end{eqnarray}
and both electrons on the right is:
\begin{eqnarray}
\mathcal{P}^{S}_\mathcal{RR} = | \langle 0 | \hat{b}_{R} \hat{b}_{R}  \hat{a}^{\dagger}_{L} \hat{a}^{\dagger}_{R} | 0 \rangle |^{2} = 0. \label{p_rr}
\label{B_RR}
\end{eqnarray}

The explicit computation of these probabilities is done in \ref{App0}. The scenario  shown in \fref{fig1} can also be interpreted as a type of two-particle interference Hong-Ou-Mandel (HOM) experiment analyzed some time ago for photons \cite{HOM}. The same scattering formalism (with the proper commutations properties for the creation and annihilation operators) has been also used to successfully analyze this type of HOM experiments for electrons \cite{Buttiker2, Buttiker3,CScho,Buttiker1}. \\

In order to properly explain the motivation and originality of the present work, we differentiate three basic steps in the computation of (two or many-particle) scattering processes. The first step is the definition of the initial state. The second is the computation of its time-evolution. The third step is providing observable results. Each one of these steps has its own conceptual and practical difficulties. In this work, we will not discuss the difficulties related to the last step, i.e. the measurement of the electrons after being scattered. We will assume a simple position measurement of electrons at the final time. The difficulties in the second step are mainly due to the many-body problem. From a computational point of view, the solution of the wave function associated to a large amount of interacting identical electrons in the high-dimensional configuration space is totally inaccessible. Approximations are mandatory. One example is the computations of the evolution of the many-body wave function through the so-called \emph{time-dependent variational principle}. This approach have been widely applied in the literature for \emph{fermionic molecular dynamics} \cite{FMD,FMD2,FMD3,FMD4}. Another example is the full configuration interaction procedure that is able to solve the many-body time evolution by an (infinite) linear combination of Slater determinants made up from one-particle basis states. A much simpler approximation is based on the concept of quasi-particles. It has been shown that a unique Slater determinant of one-particle wave functions is still able to capture most of the physics of quantum many-body problems \cite{Fagas}. The concept of one-particle wave function that still captures many-body correlations has been deduced by Oriols et al. \cite{Oriols} through the use of the conditional wave function. Strictly speaking, such quasi-particle states are no longer interpreted as a mathematical base of the system, but as a close representation of the \emph{physical} wave function.\\   

We discuss now the first step, i.e. the definition of the initial state. Let us notice that in many experiments the difficulties of the first step simply disappear because the system departs from equilibrium (i.e. the initial state is the minimum energy ground state). In other types of experiments, some controlled preparation of the initial state is possible. For example, by measuring the system at the initial time, the wave function collapses into an eigenstate of the measured eigenvalue. Unfortunately, the definition of the initial state is not so obvious in most scattering experiments. For example, in mesoscopic physics, the battery imposes far from equilibrium conditions and there is no controlled preparation of the initial state.  In principle,  one can expect some type of localization of the initial state because of the decoherence effects in the leads. On the other hand, it seems reasonable to expect that the initial quasi-particle wave packet has either positive or negative momentum (to be able to travel from the battery till the device active region). Clearly, the initial state has to be defined somewhere between two limits. The first limit is assuming that the (quasi-)electrons are described by point-like states in real space (very narrow wave packets). This limit seems quite unphysical. The other limit is describing (quasi-)electrons by point-like states in the momentum space (infinitely space extended wave packets). This second limit has been demonstrated to be very successful in the literature. The celebrated scattering probabilities presented above in equations \eref{B_LR}, \eref{B_LL} and \eref{B_RR}  fit within this second limit. However, due to the uncertainty principle, a perfectly defined momentum implies an infinitely  extended wave function in real space. Therefore, it seems reasonable to expect that the initial state is some type of intermediate state between a point-like state in real space and point-like state in momentum space (probably, quite close to the typical point-like states in momentum state, but not exactly identical to them). In the context of scattering in mesoscopic physics, similar ideas have been discussed previously by Landauer and Martin \cite{Landauer1}. Finally, let us notice that in many scenarios associating initial states to mathematical basis can be misleading. For example, in the context of Rutherford scattering, Van Boxen \emph{et al.} noticed that \lq\lq{}One must take care in calculating observables using only these basis states, because they can present misleading results when comparing to experiment.\rq\rq{}\cite{new}\\ 

With this motivation, our work analyzes the mentioned two-particle scattering experiment using  general  localized time-dependent wave packets as initial one-particle states. We compute the probability of detecting two electrons at the same side of the barrier from the anti-symmetric (Slater determinant) solution of the time-dependent Schr\"odinger equation in the configuration space \footnote{The only assumption is that this type of two-electron interference experiments can be perfectly understood from non-relativistic quantum mechanics, as most of the electron problems in chemistry and solid-state physics.}. We anticipate that in many scenarios (double and single barrier) the probability of finding two identically injected electrons at the same side of the barrier differs from zero (contrarily to what is found in expressions \eref{B_LR}, \eref{B_LL} and \eref{B_RR} for time-independent scattering eigenstates).  Our results are in qualitative agreement with the recent experiment of Bocquillon \emph{et al.} \cite{Boc}, where unexpected non-zero probabilities of detecting both electrons at the same side of the barriers were obtained. They used single-electron sources in order to ensure that two spatially localized wave packets with disjoint support were prepared at the initial time. In addition, non-zero probabilities were also found when beams of electrons were used by Liu \emph{et al.} in a similar experiment \cite{Tarucha}, Our work suggests that these non-zero probabilities are not due to experimental spurious effects \footnote{\emph{\lq\lq{}The noise is not completely suppressed, in part because of non-idealities in the beam splitter’s scattering matrix.\rq\rq{}}\cite{Tarucha}.} \footnote{\emph{\lq\lq{}The states are not perfectly identical as shown by the fact that the dip does not go to zero.\rq\rq{}}\cite{Boc}.}., but due to the fundamental wave packet nature of the electrons present in such experiments. Let us mention that the concept of quasi-particle wave packets was used by Kohn et al. \cite{kohn} some years ago when analyzing the effect of the physical borders on localized (not infinite) systems.  For the same reason, it is interesting to notice the work of Loudon \cite{Loudon} who studied the role of the exchange interaction on the scattering probabilities from a Slater-determinant of quasi-particle wave packets.\\ 

Finally, let us clarify that in time-independent computations of scattering processes, as the ones in expressions \eref{B_LR}, \eref{B_LL} and \eref{B_RR}, the difficulties of step two to determine the time evolution disappear (or equivalently the difficulties about the initial state and its evolution are packed simultaneously). There are, however, several argumentations in the literature in favor of time-dependent formalisms with wave packets to properly treat quantum transport. Vignale and Di Ventra argue that the Landauer-B\"uttiker approach is unable to capture all many-body correlations \cite{diVentra}.  The works of Kurth et al. \cite{errorDFT} and Niehaus and co-workers \cite{Niehaus} (among many others \cite{Angel,Ferry,rungePRL84,TDDFTcurrent}) show how the static density functional theory (DFT) needs to be generalized to the time-dependent DFT to properly account for non-equilibrium quantum transport. Varga showed that the nonequilibirum Green\rq{}s functions approach coupled to static DFT is not in agreement with experiments, while the time-propagation of time-dependent DFT wave packets provides better results \cite{varga}. Kon\^opka and Bokes developed a mixed basis set using stroboscopic wave packets and localized atomic orbitals to deal with interacting fermions \cite{Bokes}. Many techniques for quantum transport with wave packets have been developed by Kramer et al. \cite{kramer}. Finally, the work of Godby and co-workers show how the (Kohn-Sham) exchange-correlation potential of the time-dependent DFT can be deduced from physical quasi-particle wave packets \cite{godby}. 

After this introduction, in \sref{sec2} we define the general expressions for the probabilities of detecting two electrons. In \sref{sec3}, we generalize the previous two-electron probabilities towards many-particle probabilities. In \sref{sec4}, numerical test for typical tunneling scenarios with symmetric or non-symmetric potentials are presented. Among others, to go beyond the Fermi liquid picture of (non-interacting) electrons, we include a subsection where Coulomb interaction among the electrons in a two-particle system is included. Additionally, a single barrier potential, which is a physical system closer to the HOM experiments mentioned above, is also analyzed numerically. In all the five scenarios discussed in \sref{sec4} we obtain clear non-zero probabilities of detecting two electrons at the same side of the barrier. Conclusions are summarized in \sref{sec5}. In \ref{App0} we develop the two-particle probabilities from the scattering formalism with mono-energetic initial states. In \ref{App1} we deduce explicitly the two-particle probabilities for wave packet of arbitrary shapes. Finally, in \ref{App2}, we explicitly deduce the generalization of the previous two-particle probabilities to many-particle probabilities for tunneling scenarios with a large number of correlated electrons described by arbitrary wave packets.

\section{Two-particle probabilities}
\label{sec2}

We consider two particles injected from two different sources and impinging upon a tunneling barrier as indicated in figure~\ref{fig1}. In order to simplify the discussion, we consider electrons with identical spin orientations. Each one is individually defined in a 1D physical space. The two-particle quantum system can be defined by the (orbital) wave function $\Phi \equiv \Phi(x_1,x_2,t)$ in the 2D configuration space. Such wave function is the solution of the many-particle (non-relativistic) Schr\"odinger equation:   
\begin{equation}
i \hbar \frac{\partial \Phi}{\partial t} = \left[ - \frac{\hbar^2}{2m}\frac{\partial^2}{\partial x_1^2} - \frac{\hbar^2}{2m}\frac{\partial^2}{\partial x_2^2} + V(x_1,x_2) \right] \Phi,
\label{eq-2dexact}
\end{equation} 
where $m$ is the electron mass and $V(x_1,x_2)$ takes into account the two-particle Coulomb interaction between the electrons and also the one-particle interaction between one electron and a tunneling barrier. The exchange interaction is introduced in the shape of the initial wave function $\Phi(x_1,x_2,t_0)$. The anti-symmetrical/symmetrical (orbital) many-particle wave function for Fermions/Bosons is:
\begin{equation}
\Phi(x_1,x_2,t_0)=\frac {\phi_a(x_1,t_0)\phi_b(x_2,t_0) \mp \phi_a(x_2,t_0)\phi_b(x_1,t_0)} {\sqrt{ 2}}.  
\label{initial}
\end{equation}
The above expression can be interpreted as the determinant/permanent of a $2 \times 2$ matrix constructed from the one-particle wave function $\phi_{a}(x,t_0)$ and $\phi_b(x,t_0)$ \cite{Cohen}. Hereafter, upper/lower signs correspond to (non-relativistic) massive Fermions/Bosons. Although we mainly deal with electrons (fermions), we will also compute probabilities for  (massive) Bosons. The initial one-particle wave functions $\phi_{a}(x,t_0)$ and $\phi_b(x,t_0)$ in expression \eref{initial} are completely general. The only relevant condition for $\phi_{a}(x,t_0)$ is that its modulus square is normalizable to unity and it is totally located at the left of the barrier at time $t=t_0$. Identical conditions for $\phi_b(x,t_0)$ which is localized at the right. Additionally, according to the type of HOM experiment discussed here, both wave packets have opposite (central) momentum so that they impinge upon the barrier after a while, as depicted in figure \ref{fig1}. By construction, the time evolution of $\Phi(x_1,x_2,t)$ using equation \eref{eq-2dexact} preserves the initial norm and the initial (anti)symmetry of the wave function.\\

Let us notice why it is reasonable to expect non-zero probability for finding both electrons at the same side of the barrier. The reason is quite simple and intuitive. Pauli principle forbids two fermions being at the same position with the same state \cite{Pauli}. However, a pertinent question appears: \emph{When is reasonable the assumption that the reflected and transmitted states are exactly identical?} Certainly, both transmitted and reflected states  are identical when only one state is available in the spatial region where they coincide. This restriction on the available states is evident when the initial state has a unique (well-defined) energy $E_k$ i.e. a mono-energetic state. Then, because of the elastic nature of the interaction with the barrier (i.e. energy conservation), only one state at the right of the barrier and one at the left with the same energy $E_k$ (and the pertinent momentum going outside from the barrier) are available at the final time. Nevertheless, as stated previously, we require a superposition of mono-energetic eigenstates (i.e. a wave packet) to describe an initial state with a spatially localized support outside of the barrier region. Then, in principle, there is the possibility of different time-evolutions for the transmitted and reflected components. In such time-dependent scenarios, one can expect probabilities different from  zero (as mentioned before) as indicated in figures \ref{fig1}(c) and \ref{fig1}(d). Notice the different shapes of the reflected and transmitted wave packet in figure \ref{fig1}.

\begin{figure}[h]
\includegraphics[scale=0.43]{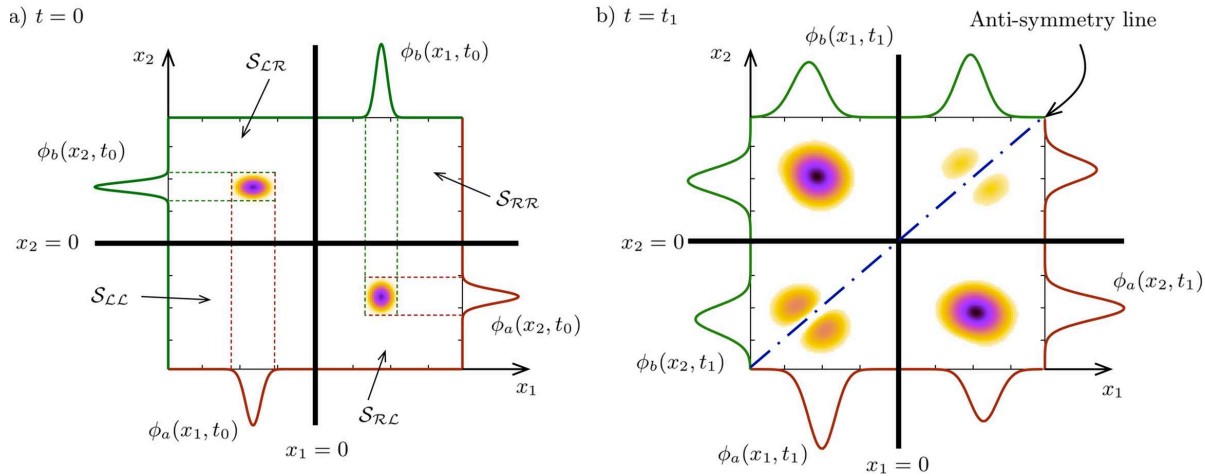}
\caption{a) Modulus square of the wave function $\Phi(x_1,x_2,t_0)$ at the initial time $t_0$ in the configuration space $\{x_1,x_2\}$. With black solid line is represented the scattering barrier. Along the axes the single particle wave packet $\phi_{a}(x,t_0)$ (red solid line) and $\phi_{b}(x,t_0)$ (green solid line) are reported for both variables $(x_1$ and $x_2)$. The dotted line visualizes how the anti-symmetrical wave function is constructed. The different region of configuration space $\mathcal{S_{LL}},\mathcal{S_{LR}},\mathcal{S_{RL}}$ and $\mathcal{S_{RR}}$ are explicitly indicated. b) Modulus square of the wave function $\Phi(x_1,x_2,t_1)$ at the final time $t_1$ (such that the interaction with the barrier is already accomplished). Along the axes $\phi_{a}(x,t_1)$ (red solid line) and $\phi_{b}(x,t_1)$ (green solid line) for both variables $\{x_1,x_2\}$ are reported . With dashed dotted blue line the anti-symmetry line for Fermions is indicated.  As asserted in the text the probabilities $\mathcal{P_{RR}} \neq \mathcal{P_{LL}} \neq 0$.}
\label{fig2}     
\end{figure}

We consider a particular time $t_1$ large enough so that the interaction with the barrier is almost finished, i.e. the probability presence inside the barrier region is negligible. Then, using Born\rq{}s rule \cite{Cohen} \footnote{The detection of the two electrons is a measurement of the wave function that implies a non-unitary evolution (not accessible from the unitary Schr\"odinger evolution). As usual, it is assumed that particle detectors provide a collapse of wave function only in those positions $\{x_1,x_2\}$ of the configuration space where the measurement is present. Notice that only reflected or transmitted components are plotted for each wave packet in figure \ref{fig1}.} in the $2D$ configuration space, $\{x_1,x_2\}$, the probability of detecting  one electron at each side of the barrier (on regions $S_{LR}$ or $S_{RL}$ of the configuration space depicted in figure \ref{fig2}(a)) at this $t=t_1$ is:
\begin{eqnarray}
\mathcal{P_{LR}} =  \int_{\mathcal{S_{LR}}} |\Phi|^2 dx_1 dx_2 +  \int_{\mathcal{S_{RL}}} |\Phi|^2 dx_1 dx_2 = 2  \int_{\mathcal{S_{LR}}} |\Phi|^2 dx_1 dx_2.
\label{P1}
\end{eqnarray}
Due to the exchange symmetry, the wave function on $\mathcal{S_{LR}}$ is identical to that on $\mathcal{S_{RL}}$, as seen in figure \ref{fig2}(a). The two integral in the left hand side of equation \eref{P1} are exactly equal, so the total contribution of finding one electron at each side of the barrier is twice one of the integrals. Equivalently, the probability of detecting the two electrons at the left of the barrier (on the region $\mathcal{S_{LL}}$ of the configuration space) is:
\begin{eqnarray}
\mathcal{P_{LL}}= \int_{\mathcal{S_{LL}}} |\Phi|^2 dx_1 dx_2.
\label{P4}
\end{eqnarray}
Finally, the probability of two electrons at the right of the barrier (on the region $\mathcal{S_{RR}}$) is:
\begin{eqnarray}
\mathcal{P_{RR}}= \int_{\mathcal{S_{RR}}} |\Phi|^2 dx_1 dx_2.
\label{P3}
\end{eqnarray}
We define $\mathcal{P_{LR}}$, $\mathcal{P_{LL}}$ and $\mathcal{P_{RR}}$ as two-particle probabilities. In figure \ref{fig2}(a) we plot the probability presence of the initial two-particle state in the 2D configuration space. According to equation \eref{initial}, the wave packet $\phi_a(x_1,t_0)\phi_b(x_2,t_0)$ has its support on $\mathcal{S_{LR}}$, while the wave packet $\phi_a(x_2,t_0)\phi_b(x_1,t_0)$ on $\mathcal{S_{RL}}$. There is no initial probability presence in the other regions. The first relevant issue seen on the regions $\mathcal{S_{LL}}$ and $\mathcal{S_{RR}}$ of figure \ref{fig2}(b) is that $\mathcal{P_{LL}} \neq 0$ and $\mathcal{P_{RR}} \neq 0$. We explain the reason of these non-zero probabilities in next section. In general, let us mention that there is no reason to expect that the probability of detecting two electrons at the left of the barrier is equal to the probability of detecting them at the right,  $\mathcal{P_{RR}} \neq \mathcal{P_{LL}}$  as seen in figure \ref{fig2}(b). 

\begin{figure}[h]
\includegraphics[scale=0.45]{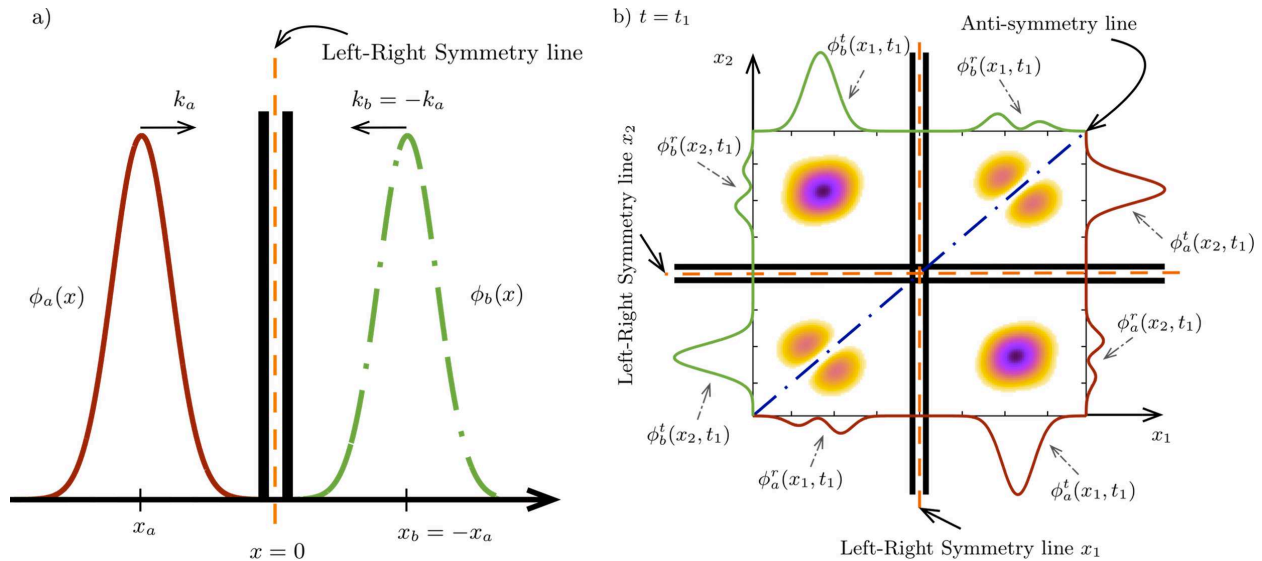}
\caption{a) Schematic representation of the initial wave packets in the physical space under the conditions (i), (ii) and (iii). With black solid line the double barrier structure is depicted. With orange dashed line the Left-Right symmetry of the problem is depicted. With red solid line wave packet $\phi_{a}(x)$ centered in $x_a$ and with momentum $k_a$ is depicted. With green dashed dotted line the wave packet $\phi_{b}(x)$ centered in $x_b = -x_a$ and momentum $k_b=-k_a$ is reported . b) Modulus square of the wave function $\Phi(x_1,x_2,t_1)$ at the final time $t_1$ at the configuration points $\{x_1,x_2\}$. With dashed dotted blue line the anti-symmetry line for Fermions is indicated and with orange dashed lines the Left-Right symmetry for each degree of freedom ($x_1$ and $x_2$) is reported. Along the axes, the modulus square of the $\phi_{a}$ (red line) and $\phi_{b}$ (green line) wave functions are plotted for each degree of freedom ($x_1$ and $x_2$). The upperindices $r$ or $t$ indicate reflected or transmitted components, respectively.}
\label{fig3}    
\end{figure}

To certify the unavoidable fundamental (not spurious) origin of the non-zero probabilities for $\mathcal{P_{LL}}$ and $\mathcal{P_{RR}}$, hereafter, we consider exactly the same idealized conditions used in Refs. \cite{Buttiker2,Buttiker3,CScho, Buttiker4} when they discuss the two-particle probabilities. We take the two wave packets $\phi_{a}(x,t_0)$ and $\phi_{b}(x,t_0)$ as identical as possible. In particular, we impose the following three conditions:
\begin{itemize}
\item Condition (i): A separable potential $V(x_1,x_2)$ in equation \eref{eq-2dexact} without Coulomb interaction: 
\begin{equation}
V(x_1,x_2)=V_B(x_1)+V_B(x_2), 
\label{potsep}
\end{equation}
where $V_B(x)$ is the symmetrical potential energy of a tunneling barrier, i.e. $V_B(x)=V_B(-x)$, with $x=0$ at the center of the barrier region. See figure \ref{fig3}(a). 
\item Condition (ii): All parameters of the initial wave packet $a$ and $b$ are identical, except for the initial central momentums which accomplishes $k_b=-k_a$ and central positions $x_b=-x_a$. See figure \ref{fig3}(a).
\item Condition (iii): Electrons are injected exactly at the same time. 
\end{itemize}
Because of these conditions, as discussed in \ref{App1}, the two initial wave packets are defined with (almost) identical parameters. In particular, we have $g_a(k)=g_b(-k)$ where $g_{a}(k)=\braket { \phi_{a}(x,t_0)} { \psi_k(x)}$ is the complex value that weights the superposition of the scattering states to build the wave packet $\phi_{a}(x,t_0)$. See expression \eref{paquetesuperposition} in \ref{App1}. Identical definition for $g_b(k)$.  

Under these conditions, we can anticipate the evolution of $\Phi(x_1,x_2,t)$ and also the origin of the non-zero probabilities for arbitrary wave packets. We consider the initial (anti-symmetrical) wave function of two electrons $\Phi(x_1,x_2,0)$ defined by equation \eref{initial}. Since the time-evolution of Schr\"odinger equation satisfies the superposition principle, we can discuss the time-evolution of $\phi_a(x_1,t_0)\phi_b(x_2,t_0)$ and $\phi_a(x_2,t_0)\phi_b(x_1,t_0)$ independently. Then, since we are dealing with a separable Hamiltonian, the evolution of $\phi_a(x,t)$ and  $\phi_b(x,t)$ can be computed from two simpler single particle Schr\"odinger equations. At a time $t=t_1$, after the interaction with the barrier,  each wave packet splits into two (non-overlapping) components: 
\begin{eqnarray}
\phi_{a}(x,t_1)=\phi_{a}^r(x,t_1)+\phi_{a}^t(x,t_1), \label{separa1}\\
\phi_{b}(x,t_1)=\phi_{b}^r(x,t_1)+\phi_{b}^t(x,t_1), \label{separa2}
\end{eqnarray}
where the upperindices $r$ and $t$ refer to the reflected and transmitted component of each wave packet ($\phi_a$ and $\phi_b$), respectively. Then, the two particle wave function in the region of the configuration space $\mathcal{S_{LL}}$ at $t=t_1$ is:
\begin{equation}
\Phi(x_1,x_2,t_1)\big|_{\mathcal{S}_{\mathcal{LL}}}=\frac {\phi^r_a(x_1,t_1)\phi^t_b(x_2,t_1) - \phi^r_a(x_2,t_1)\phi^t_b(x_1,t_1)} {\sqrt{ 2}}.
\label{exchangeexample}
\end{equation}

Let us notice that the region $\mathcal{S_{LL}}$ was initially empty of probability, as seen in figure \ref{fig2}(a). The initial wave packet $\phi_a(x_1,t_0)$ on $\mathcal{S}_{\mathcal{LR}}$ (which is identical to the one plotted in figure \ref{fig2}(a)) evolves into the part $\phi^r_a(x_1,t_1)$ on $\mathcal{S}_{\mathcal{LL}}$ in figure \ref{fig3}(b).  Equivalently, the initial wave packet $\phi_b(x_2,t_0)$ in figure \ref{fig2}(a) evolves into the part $\phi^t_b(x_2,t_1)$ on $\mathcal{S}_{\mathcal{LL}}$ in figure \ref{fig3}(b).  Identical explanations for the presence of $\phi^t_b(x_1,t_1)$ and $\phi^r_a(x_2,t_1)$ on $\mathcal{S}_{\mathcal{LL}}$.  Clearly, since $\mathcal{P}_{\mathcal{LL}}$ in equation \eref{P4} is computed from an integral of non-negative real numbers,  the requirement for obtaining the result $\mathcal{P}_{\mathcal{LL}}=0$ in equation \eref{exchangeexample} is that  $\phi^r_a(x_1,t_1)\phi^t_b(x_2,t_1) = \phi^r_a(x_2,t_1)\phi^t_b(x_1,t_1)$ at all positions $\{x_1,x_2\} \in \mathcal{S}_{\mathcal{LL}}$  \footnote{By construction, only in the configuration space points $\{x_1,x_2\} \in \mathcal{S}_{\mathcal{LL}}$ such that $x_1 = x_2$, the wave function is always strictly zero (see anti-symmetry line in the figure \ref{fig3} b).}. This last condition can only be obtained when $\phi^{t}_{b}(x,t_1) = \phi^{r}_{a}(x,t_1)$ and  $\phi^{t}_{a}(x,t_1) = \phi^{r}_{b}(x,t_1)$. On the contrary, if the transmitted and reflected wave packet components differ, i.e. if the time-evolution giving the transmitted component $\phi^{t}_{a}(x,t_1)$ is different from $\phi^{r}_{b}(x,t_1)$, then we get  $\Phi(x_1,x_2,t_1)\neq 0$, which implies  $\mathcal{P}_{\mathcal{LL}} \neq 0$. Analogous consideration can be done for the configuration space region $\mathcal{S}_{\mathcal{RR}}$. 

After discussing the origin of the non-zero probabilities, we present a technical question that we will test numerically later.  The conditions (i), (ii) and (iii) impose an additional symmetry on the problem. Apart from the intrinsic anti-symmetry of the wave function implicit in equation \eref{initial}, there is an additional Left-Right symmetry. This means that, being $x=0$ the center of the barrier region as depicted in figure \ref{fig3}(a), the wave function under the separable Hamiltonians of expression \eref{potsep} has to satisfy $\Phi(x_1,x_2,t) = -\Phi(-x_1,-x_2,t)$ at all times. This additional symmetry implies that the probability of detecting two electrons on the left is exactly equal to detect them on the right, i.e. $\mathcal{P}_{\mathcal{LL}} = \mathcal{P}_{\mathcal{RR}}$ as depicted in figure \ref{fig3}(b). However, let us notice that, in general, when conditions (i), (ii) and (iii) are not satisfied, we have  $\mathcal{P}_{\mathcal{LL}} \neq \mathcal{P}_{\mathcal{RR}}$ as depicted in the preceding figure \ref{fig2}(b).

The exact values of $\mathcal{P_{LR}}$, $\mathcal{P_{LL}}$ and $\mathcal{P_{RR}}$ depend on the effective overlapping between $\phi^{t}_{a}(x,t_1)$ and $\phi^{r}_{b}(x,t_1)$. In \ref{App1} we develop analytical calculations of the range of values that the probabilities \eref{P1}-\eref{P3} can take when conditions (i), (ii) and (iii) are assumed. When reflected and transmitted wave packets are identical as indicated in expression \eref{condition1}, expressions \eref{P1}-\eref{P3} can be rewritten as:
\begin{eqnarray}
\mathcal{P}^{M}_{\mathcal{LL}}=\mathcal{P}^{M}_{\mathcal{RR}}=R T \mp R T,\;\;\;\;\;\;
\label{P4M}\\
\mathcal{P}^{M}_{\mathcal{LR}}=(R \pm  T)^2,\;\;\;
\label{P1M}
\end{eqnarray}
which corresponds to the well-known result $\mathcal{P}^{M}_{\mathcal{LL}}=\mathcal{P}^{S}_{\mathcal{LL}}=0$, $\mathcal{P}^{M}_{\mathcal{RR}}=\mathcal{P}^{S}_{\mathcal{RR}}=0$ and $\mathcal{P}^{M}_{\mathcal{LR}}=\mathcal{P}^{S}_{\mathcal{LR}}=1$ mentioned in equations (\ref{B_LR}), (\ref{B_LL}) and (\ref{B_RR}) for fermions. Additionally, we have $\mathcal{P}^{M}_{\mathcal{LL}}=\mathcal{P}^{M}_{\mathcal{RR}}=2 R T$ and $\mathcal{P}^{M}_{\mathcal{LR}}=(R-T)^2$ for Bosons. Let us notice that the sum of the three probabilities is equal to one (for Fermions or Bosons) because we deal with a unitary evolution. We use the upperindex $M$ denoting that the overlapping between the transmitted and reflected components is maximum. In summary, we have tested that our general definitions of the two-particle probabilities in equations (\ref{P1})-(\ref{P3}) exactly reproduce, as a particular example, the results found in the literature for scattering states in Refs. \cite{Buttiker2,Buttiker3,CScho,Landauer1,Landauer2}.

For other scenarios, for example a double barriers with wave packets with resonant energies, we show in \ref{App1} that the transmitted and reflected components become orthogonal. Then,  the probabilities (\ref{P1}) - (\ref{P3}) in this type of experiments at resonances can be written as:
\begin{eqnarray}
\mathcal{P}^{m}_{\mathcal{LL}}=\mathcal{P}^{m}_{\mathcal{RR}}=R T,
\label{P4m}\\
\mathcal{P}^{m}_{\mathcal{LR}}=R^2 + T^2.
\label{P1m}
\end{eqnarray}
where the upperindex $m$ here indicates that the overlapping between transmitted and reflected components is zero (minimum). Again, the sum of the probabilities is one because of the unitary evolution. These last probabilities $\mathcal{P}^{m}_{\mathcal{LL}},\mathcal{P}^{m}_{\mathcal{RR}}$ and $\mathcal{P}^{m}_{\mathcal{LR}}$ show no difference between Fermions or Bosons. In fact, these results are identical to the probability of distinguishable particles. In conclusion, even with both electrons at the same position at the same time, the Pauli principle has no effect in these HOM scenarios because the wave nature of electrons is described by different (orthogonal) wave functions. We emphasize that, in general, the two-particle probabilities in equations (\ref{P1})-(\ref{P3}) can take any value between the limits imposed by expressions \eref{P4M}-\eref{P1M} and \eref{P4m}-\eref{P1m}. We emphasize that all the previous results are valid for any shape of quasi-particle wave packets.  

\section{N-particle probabilities}
\label{sec3}

We generalize the previous two-particle scattering probabilities towards many-particle scattering probabilities. We consider a scenario similar to the situation depicted in \fref{fig1} with one electron at each side of the barrier, at the same distance. Now, we add $N-2$ electrons distributed arbitrarily at both sides.  The anti-symmetric normalized wave function of the $N$-electrons is the following:
\begin{equation}
\Phi_N=\frac{\phi_N}{\sqrt{(\int^\infty_{-\infty} dx_1 ... \int^\infty_{-\infty} dx_N |\phi_N|^2)}} ,
\label{phinorm}
\end{equation}
where $\phi_N$ is equal to
\begin{equation}
\phi_N=\frac{1}{\sqrt{N!}}\sum_{p\in S_N}\left(\prod_{l=1}^N\psi_l(x_{p(l)},t) \right)f(p) ,
\label{wpN}
\end{equation}
with $S_N$ is the group of the $N!$ permutations $p$ on the set of $N$ electron positions ${x_1,x_2, ... , x_N}$ and $f(p)=\pm 1$ is equal to the sign of the permutations. For simplicity, we avoid again the discussion about spin. The shape of each individual wave packet $\psi_l(x,t)$ in equation \eref{wpN} is arbitrary satisfying the same conditions mentioned in \sref{sec2}. In particular, they are normalized to unity $\int_{-\infty}^\infty \psi_l(x,t)dx=1$ and initially localized at one side of the barrier. The denominator in equation \eref{phinorm} appears to properly normalize the N-electron wave function if some of the single-particle wave-packets overlap. 

We are interested in computing the probability of finding the $N$ electrons at one side of the barrier, for example, the right one. Such probability can be computed as: 
\begin{equation}
\mathcal{P_{R^N}}=\int^\infty_0 dx_1 ... \int^\infty_0 dx_N |\Phi_N|^2,
\label{qq}
\end{equation}
with $\mathcal{P_{R^N}}\equiv\mathcal{P_{R R R ..... R}}$ meaning the probability of finding the $N$ electrons at the right of the barrier. Other probabilities mixing left and right detection can also be identically defined. We will only focus on the type of probability of \eref{qq} which is the relevant one when discussing the unexpected non-zero probabilities.  For practical purposes, it can be straightforwardly demonstrated that the probability $\mathcal{P_{R^N}}$ can be computed as:

\begin{equation}
\mathcal{P_{R^N}}=\sum_{p \in S_N}f(p) \prod_{l=1}^N p_R(l,p(l)).
\label{qqpractical}
\end{equation}

The complex value $p_R(l,j)$  is a correlation function between single-particle wave packets, defined as:

\begin{eqnarray}
p_R(l,j)&=&\frac{1}{\mathcal{P_N}^{\frac{1}{N}}}\int^\infty_0\psi_l^*(x,t) \psi_j(x,t)dx,
\label{q}
\end{eqnarray}

where $\mathcal{P_N}$ is defined as the denominator of equation \eref{phinorm}.
This correlation function is negligible if the initial wave packets do not overlap. It is straightforward to notice that $p_R(l,j)=p_R(j,l)^*$. It is interesting to realize that the probability in equation \eref{qq} is just the determinant of the following matrix, $\mathcal{P_{R^N}} = det(M_{R^N})$

\begin{eqnarray}
\left(\begin{array}{cccc}  D_1 & p_R(1,2) & ... & p_R(1,N) \\ p_R(2,1) & D_2 & ... & p_R(2,N)\\ ... & ... & ... & ... \\ p_R(N,1) & p_R(N,2) & ... & D_N  
  \end{array} \right).
 \label{matrixQQ}
\end{eqnarray} 
The terms $D_i$ corresponds to the reflection probability of the i-wave packet $R_i$ if it was at the right side at the initial time or to the transmission probability of the i-wave packet $T_i$ if it was at the left. The consideration of the probabilities for Bosons will imply dealing with the permanent of $M_{R^N}$. 

We can extract, two limiting cases for the probability $\mathcal{P_{R^N}} = det(M_{R^N})$. When we have two individual wave packets $j$ and $k$ identical among the set of $N$ electrons, $p_R(l,j)=p_R(l,k)$ for any $l$, we get:
\begin{equation}
\mathcal{P^{M}_{R^N}}=0.
\label{qq_overlapping}
\end{equation}
because the determinant of a matrix with two columns (rows) identical is zero. From what we learnt in \sref{sec2}, these zero probabilities appear, for example, when one reflected and one transmitted wave packets are very close to mono energetic states with a  unique energy. This results in equation \eref{qq_overlapping} is a generalization of expression \eref{P4M} for two electrons. On the contrary, it is quite simple to realize that in the case where there is no overlapping between any two different individual wave packets, $p_R(l,j)=0$ for $l \neq j$, we get:
\begin{equation}
\mathcal{P}^{m}_{R^N}=D_1\cdot D_2 \cdot ..... \cdot D_N,
\label{qq_nooverlapping}
\end{equation}
because the determinant of a diagonal matrix is just the product of its diagonal elements. This result is just a many-particle generalization of expression \eref{P4m} for electrons.  From what we learn in \sref{sec2}, this result \eref{qq_nooverlapping} appears, even when wave packets share identical spatial regions, if they are orthogonal. In conclusion, the two limiting cases discussed in \sref{sec2} are identically found for many-particle scenarios. \\

Fortunately, some useful properties for $\mathcal{P_{R^N}} = det(M_{R^N})$ can be deduced without fixing any individual property of the wave packets. In the \ref{App2}, we show that for any set of $N$ (quasi-particle) wave packets,  when we consider a subset of $K<N$ electrons among them, the probabilities of the two sets must satisfy the condition:  
\begin{equation}
\mathcal {P_{R^N}}  \leqslant  {D_N\cdot D_{N-1}\cdot ...\cdot D_{K+1}} \; \cdot  \mathcal{P_{R^K}}
\label{proof2}
\end{equation}
The term $\mathcal{P_{R^K}}$ is the probability of finding the set of electrons $\{1,2,...,K\}$ at the right of the barrier when the exchange interaction among them is considered. On the other hand,  ${D_N\cdot D_{N-1}\cdot ...\cdot D_{K+1}}$ in equation \eref{proof2} is just the probability of finding the set of electrons $\{N,N-1,...K+1\}$ at the right of the barrier when exchange interaction among them is not relevant. From this result, one can expect that the non-zero probabilities of simultaneous electrons at the right will tend to zero value when more and more electrons with similar central positions and momentums are considered in a limited region of the (position-momentum) phase-space. On the contrary, if such additional electrons are associated to wave packets with different central position or momentum (or partially orthogonal), their inclusion is not so irrelevant.  The phase-space density of electrons plays a crucial role in determining which expression \eref{qq_nooverlapping} or \eref{qq_overlapping} is relevant in each scenario. The discussion of the probability $\mathcal{P_{R^N}} = det(M_{R^N})$ as a function of the phase-space density will be discussed numerically in \sref{sec3.2}. 

\section{Numerical results}
\label{sec4}

Next, we confirm numerically the predicted non-zero probabilities in different and general scenarios. Our procedure for the numerical computation of the probabilities $\mathcal{P}_{\mathcal{LR}}$, $\mathcal{P}_{\mathcal{LL}}$ and $\mathcal{P}_{\mathcal{RR}}$ is the following.  First, we time-evolve an (anti-symmetrical) initial state, defined by expression \eref{initial}, with the Schr\"odinger equation in the configuration space. Second, at the final time $t_1$, we compute the different probabilities \eref{P1}-\eref{P3} from the modulus square of the wave function through Born's rule, without any approximation. We will follow an equivalent procedure to compute many-particle probabilities. 

\subsection{Two-particle scenario with a separable and symmetrical double barrier potential}
\label{sec31}

\begin{figure}[h]
\includegraphics[scale=0.40]{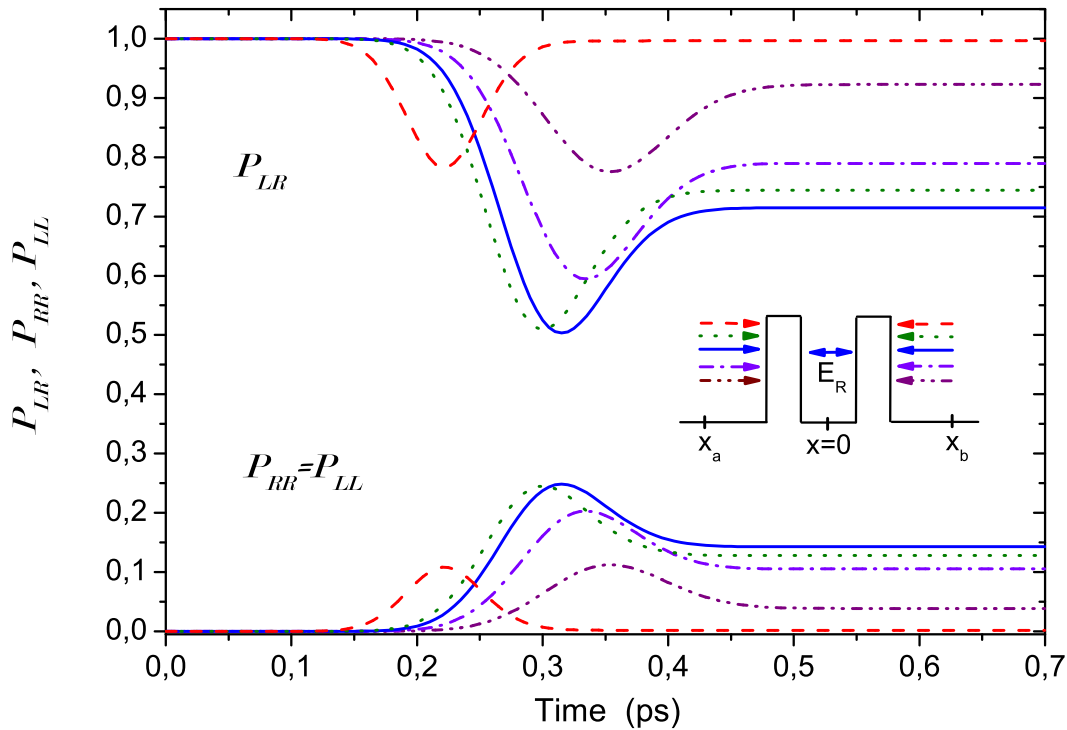}
\caption{Time evolution of $\mathcal{P}_{\mathcal{LR}}$ (upper lines) and $\mathcal{P}_{\mathcal{LL}}=\mathcal{P}_{\mathcal{RR}}$ (lower lines) from $\Phi(x_1,x_2,t)$ built by two initial wave packets located at $x_a=-175$ nm and $x_b=175$ nm with opposite momentums and equal spatial dispersions $\sigma_a=\sigma_b=35\;nm$. The energies are $E_a=E_b=0.12\;eV$ (red dashed line),   $E_a=E_b=0.075\;eV$ (green dot line),  $E_a=E_b=E_R=0.069\;eV$ (blue solid line),     $E_a=E_b=0.06\;eV$ (dash dot violet line) and   $E_a=E_b=0.05\;eV$ (dash dot dot purple). The inset shows the potential profile.}
\label{fig4}     
\end{figure}

We consider the double barrier drawn in figure \ref{fig3}(a) and also in the inset of figure \ref{fig4}. The potential profile is built by two barriers of 0.4 eV of height and 0.8 nm of width between a quantum well of 5.6 nm. This potential profile has Left-Right symmetry. The $x=0$ is situated at the center of the quantum well. The (effective) mass of the electrons ($m$) is 0.067 times the free electron mass. The first resonant energy of such structure is $E_R=0.069\;eV$.  At the initial time $t_0$, the initial state is defined for numerical convenience by two Gaussian wave packets (other choices are possible \cite{newnew}), $\phi_a(x,t_0)$  and $\phi_b(x,t_0)$ \cite{Cohen} whose spatial support is located at the left and right of the barrier, respectively. Let us notice that such Gaussian wave packets have point-localized or fully-extended mono-energetic states as two limiting cases. Both wave packets have the same central energy $E_a=E_b$, but opposite central wave vectors $k_b=-k_a$ and central positions  $x_a=-x_b$. In figure ~\ref{fig4} the time evolutions of expressions (\ref{P1})-(\ref{P3}) are depicted. First, we see that for a wave packet whose energy is far from the resonant energy $E_R$, we obtain  $\mathcal{P}^{S}_{\mathcal{LR}} \equiv \mathcal{P}^{M}_{\mathcal{LR}}=1$, $\mathcal{P}^{S}_{\mathcal{LL}}\equiv\mathcal{P}^{M}_{\mathcal{LL}}=0$ and $\mathcal{P}^{S}_{\mathcal{RR}}\equiv\mathcal{P}^{M}_{\mathcal{RR}}=0$, at $t_1=0.7\;ps$, as predicted by expressions \eref{B_LR}-\eref{B_RR} and also by  \eref{P4M}-\eref{P1M}. However, for the resonant energy $E_a=E_b=E_R$ we get the results $\mathcal{P}_{\mathcal{LR}}  \equiv \mathcal{P}^{m}_{\mathcal{LR}}=1-2RT$, $\mathcal{P}_{\mathcal{LL}} \equiv \mathcal{P}^{m}_{\mathcal{LL}}=RT$ and $\mathcal{P}_{\mathcal{RR}} \equiv \mathcal{P}^{m}_{\mathcal{RR}}=RT$ that correspond to the values of indistinguishable particles predicted by expressions \eref{P4m}-\eref{P1m}. To test these last expressions numerically, we notice that this potential profile and wave packets give $T=0.806$ and $R=0.194$, where $R$ and $T$ are the single particle reflection and transmission coefficients. As explained (see \ref{App1}), the latter set of probabilities correspond to a scenario in which the transmitted and reflected components are orthogonal. In other words, the transmitted wave packet is basically built by a superposition of resonant scattering states, while the reflected one by mainly non-resonant scattering states. 

\begin{figure}[h]
\includegraphics[scale=0.40]{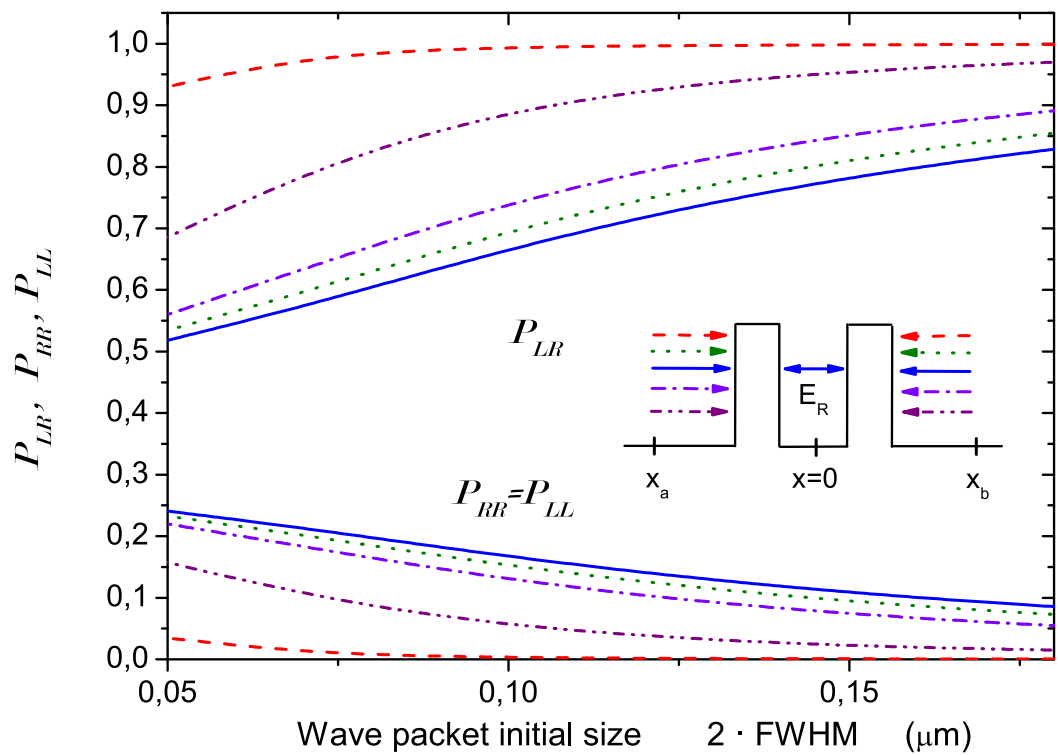}
\caption{The probabilities of $\mathcal{P}_{\mathcal{LR}}$ (upper lines)  and $\mathcal{P}_{\mathcal{LL}}=\mathcal{P}_{\mathcal{RR}}$ (lower lines) from  $\Phi(x_1,x_2,t_1)$ at time $t_1=0.7$ ps with the same initial wave packets and energies of figure~\ref{fig4} but with variable spatial dispersion $\sigma_x=\sigma_a=\sigma_b$. The inset shows the potential profile.}
\label{fig5}    
\end{figure}

As mentioned in the introduction, dealing with the time-dependent Schr\"odinger equation implies that the results depend also on the initial wave packet shape. In figure~\ref{fig5}, we study the dependence of the two-particle probabilities of figure~\ref{fig4} on the \emph{size} of the initial wave packet. We define the size of the initial wave packet as the double of the full width at half maximum (FWHM) of the probability presence of the Gaussian wave packet at $t=t_0$. Such size can be related with the spatial dispersion  $\sigma_x$ of the initial wave packet from   $2 \times  FWHM=4 \sqrt{\ln(2)} \sigma_x$. In the limit of $\sigma_x \rightarrow \infty$, a wave packet approaches to a scattering state. 

The maximum wave packet dimensions considered in figure~\ref{fig5} are much larger than typical reservoir sizes in quantum transport with semiconductors \cite{sizes} and we still clearly see $\mathcal{P}_{\mathcal{LL}}=\mathcal{P}_{\mathcal{RR}}\neq 0$. In addition, if we consider barriers much higher than 0.4 eV, the resonance becomes much sharper and wave packets with $\sigma_x \approx 1\; \mu m$ still show $\mathcal{P}_{\mathcal{LL}}=\mathcal{P}_{\mathcal{RR}}\neq 0$.

\subsection{Many-particle scenario with a separable and symmetrical double barrier potential}
\label{sec3.2}

\begin{figure}[h!]
\centering
\includegraphics[scale=0.5]{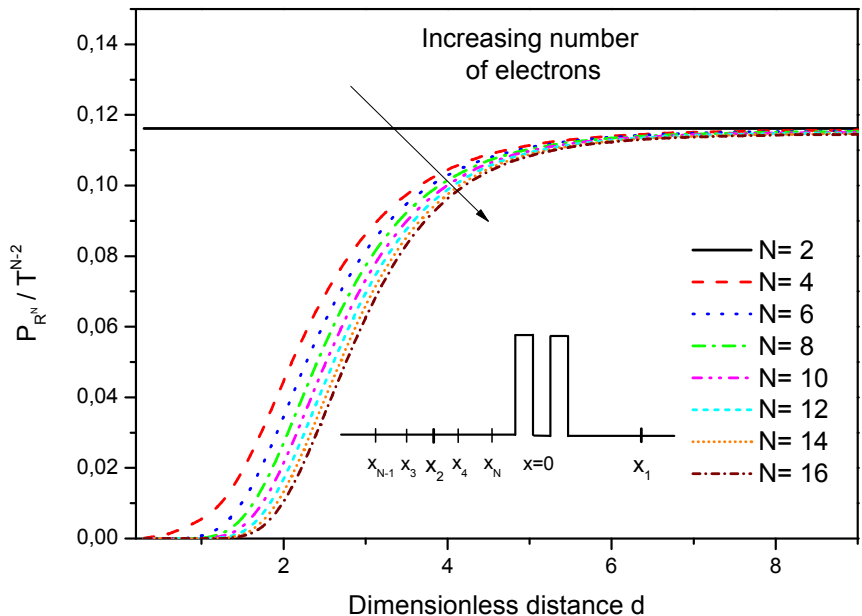}
\caption{Probability of finding all electrons at the right side of the potential barriers $\mathcal {P_{R^N}}/(T^{N-2}) $ depending on the phase-space density quantified by the dimensionless distance $d$ for different number of electrons (in the inset we see the initial configuration). At high phase-space density (small distances $d$), the probability is almost zero. It is relevant to remark that this limit is obtained for larger values of $d$ when the number of electrons is higher. At low phase-space density, when the distance among different wave packets is large, the probability of finding all electrons at the right of the barrier is still non-zero independently of the number of electrons, achieving the result of two-particle scattering obtained in \sref{sec31} as a limit.}
\label{fig13}    
\end{figure}

We discuss if the consideration of more than two electrons could invalidate the behavior of these unexpected non-zero probabilities. We study numerically the role of the phase-density of electrons in the probabilities $\mathcal {P_{R^N}}$ discussed in \sref{sec3}. We define the dimensionless distance between two wave packets in the phase-space as:

\begin{equation}
d=\frac{(k_i-k_j)^2}{2\sigma_k^2}+\frac{(x_i-x_j)^2}{2\sigma_x^2},
\label{distance}
\end{equation}

where $i$ and $j$ are two consecutive electrons and $k_{i/j}$ and $x_{i/j}$ represent the initial central momentum and initial central position of a wave packet, respectively. A small value of $d \ll 1$ means large position (or momentum) overlapping between wave packets, while a large value $d \gg 1$ means no momentum or position overlapping. The phase-space measure $d$ in equation \eref{distance} is analogous to the one used in Refs. \cite{FMD,FMD5} in the context of \emph{fermionic molecular dynamics} mentioned in the introduction. Within this theory it has also been studied the consequences of the Pauli principle for scattering of localized (Gaussian) wave packets \cite{FMD,FMD5}. The procedure to build the results shown in \fref{fig13} is to fix one initial wave packet $\psi_1(x_{1},t)$ located at the right side (in this case $D_1=R$) and then vary the number of initial wave packets at the left side, as shown in the inset of \fref{fig13}. For $N=2$ (black solid curve),  a second wave packet $\psi_2(x_{2},t)$ is initially fixed at the left with $D_2=T$. For $N=4$, red dashed curve, we add two more electrons (one at the left side and another at the right of the second electron at the same distance $d$). Both, with $D_3=T$ and $D_4=T$. Finally, for $N=16$ brown short dashed dotted curve, we have initially one electron at the right and $15$ at the left. After a proper time evolution of the $N$-particle wave function, the probability in equation \eref{qq} is computed. We plot in \fref{fig13} the results $\mathcal {P_{R^N}}/(T^{N-2})$ as a function of the distance $d$ for different number $N$ of electrons. We divided the $\mathcal {P_{R^N}}$ probability by $T^{N-2}$ in order to be able to  compare probabilities with different number of electrons. Technical details of the computation are explained in \ref{App2}.\\

We see in \fref{fig13} that the probability of each curve decreases, reaching the zero value as a limit when small values of $d$ are considered (high phase-space density). For small $d$ values, the overlapping among different wave packets is increased. If the transmitted electron avoids somehow the (position or momentum) overlapping with the first reflected electron, it unavoidably overlaps with some others. This limit of zero probability is achieved for moderate values $d$ when more and more electrons are considered. As a consequence, in agreement with the fluctuation-dissipation theorem \cite{Nyquist,Johnson}, at high density (low temperature) no noise is present in the many-particle tunneling process.  This is the well-known result found in the literature for scattering states  \cite{Buttiker2,Buttiker3,CScho,Landauer1,Landauer2}. However, for large distances $d$, the same value of the probability of two particles studied in \sref{sec31} is obtained independently of the number $N$ of electrons that we are considering. Although we have studied numerically a very particular shape and phase-space configuration of the wave packets, it is obvious that there are experimental windows (high temperatures to achieve low phase-space densities of wave packets) where the non-zero probabilities discussed here are also clearly accessible in typical (many-particle) electronic devices. 

\subsection{Two-particle scenario with a separable and non-symmetrical double barrier potential}
\label{sec32}

\begin{figure}[h]
\includegraphics[scale=0.40]{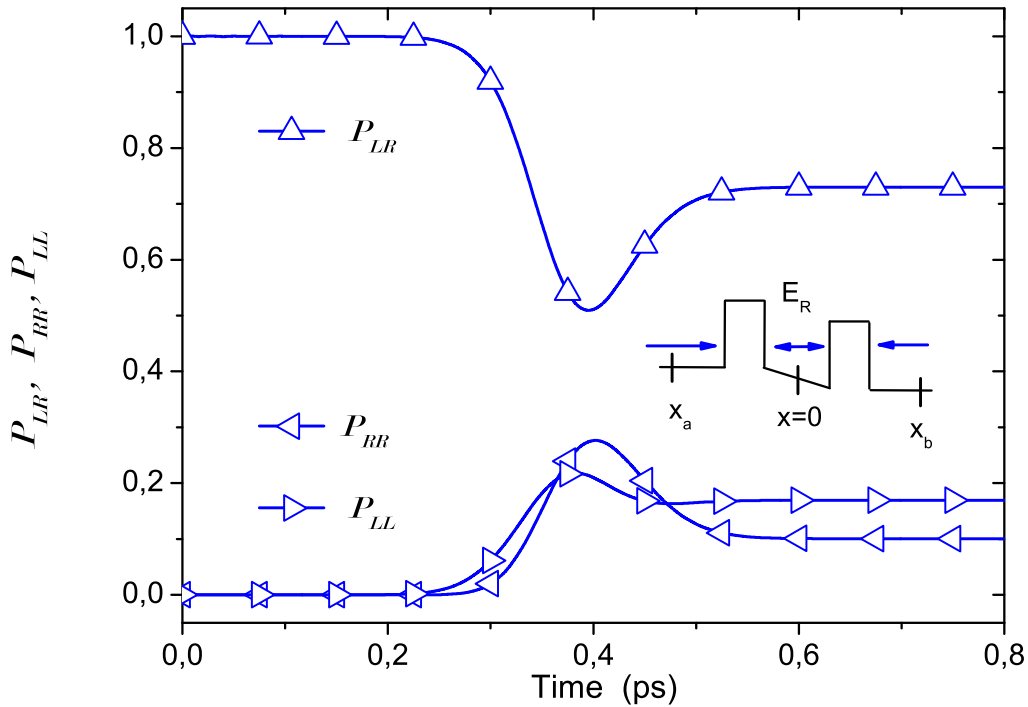}
\caption{Time evolution of $\mathcal{P}_{\mathcal{LR}}$, $\mathcal{P}_{\mathcal{LL}}$ and $\mathcal{P}_{\mathcal{RR}}$ (solid blue lines) for $\Phi(x_1,x_2,t)$ built from two initial wave packets located at $x_a=-175$ nm and $x_b=257$ nm with  equal spatial dispersions $\sigma_a=\sigma_b=35\;nm$. The inset shows the potential profile. }
\label{fig6}     
\end{figure}

Here we analyze which is the role of symmetry of the potential in the computation of non-zero probabilities. In typical mesoscopic systems, the potential profile is not Left-Right symmetrical. For example, when an external battery is included. It implies an asymmetric potential profile, as indicated in the inset of figure~\ref{fig6}. Then, the conditions (i), (ii) and (iii) are not applicable and the two-particle probabilities (\ref{P1})-(\ref{P3}) present an even richer phenomenology.  

We consider the same scenario studied in figure \ref{fig4} with an applied bias of $0.05\;V$ (see the inset of figure~\ref{fig6}). The kinetic energy of the $a$-wave packet is $E_a=E_R=0.043\;eV$ equal to the new resonant energy. The kinetic energy of the $b$-wave packet is, $E_b=E_R= (0.043+0.05)\;eV$. Different initial positions are selected to ensure that the wave packets coincide in the barrier region, at the time $x_a/v_a^c=x_b/v_b^c$, with the initial central velocity $v_a^c=\hbar k_a^c/m=4.75 \; 10^5\;m/s$ and $v_b^c=-4.88\;10^5\;m/s$. In any case, at time $t_1=0.8\;ps$, we see a rich phenomenology for the two-particle  probabilities with $\mathcal{P}_{\mathcal{LL}} \neq \mathcal{P}_{\mathcal{RR}} \neq 0$. As mentioned, the additional Left-Right symmetry is not present.

\subsection{Two-particle scenario with non-separable double barrier potential}
\label{sec4d}

\begin{figure}[h]
\includegraphics[scale=0.35]{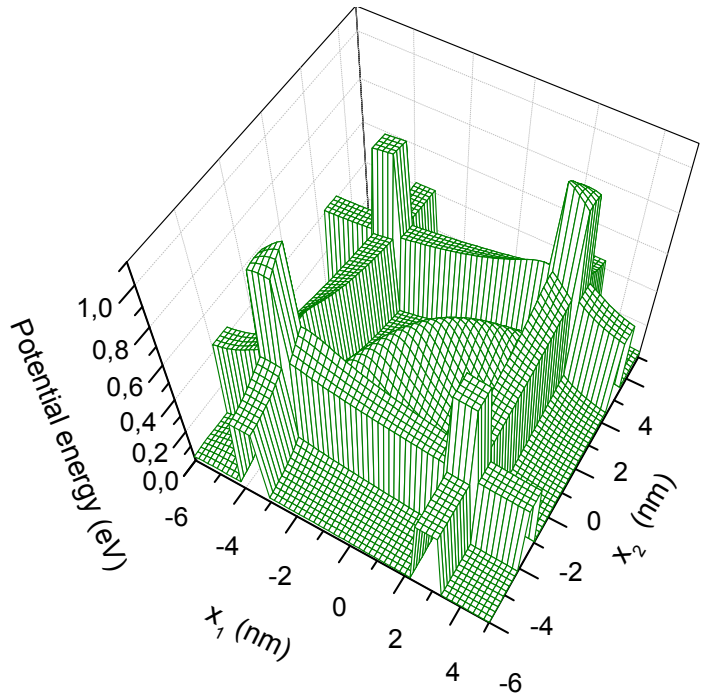}
\caption{Potential $V(x_1,x_2)$  in the configuration space $\{x_1,x_2\}$ with Coulomb interaction in a double barrier when C=5.} 
\label{fig7}    
\end{figure}

At this point, we analyze if the assumption of quasi-particle is a fundamental issue in the non-zero probabilities. We go beyond the Fermi liquid theory and compute the Coulomb interaction among two electrons in the type of HOM configuration considered here. In this subsection, we consider wave functions $\Phi(x_1,x_2,t)$ solutions of equation \eref{eq-2dexact} with the same initial expression \eref{initial} but with a non-separable potential: 
\begin{equation}
V(x_1,x_2)=V_B(x_1)+V_B(x_2) +C \cdot V_C(x_1,x_2),
\label{potnosep}
\end{equation}
being $V_C(x_1,x_2)$ the Coulomb interaction between electrons. The constant $C$ takes into account the strength of the interacting Hamiltonian (i.e. $C=0$ means separable Hamiltonian). We use the explicit expression: 
\begin{equation}
V_C(x_1,x_2) = \frac{q^2}{4 \pi \epsilon_r \epsilon_0} \frac{1}{\sqrt{(x_1-x_2)^2+a_C^2}} f(x_1,x_2),
\label{coulpot}
\end{equation}
with $\epsilon_r=11.6$ and $\epsilon_0$ is the free space dielectric constant. To avoid numerical irrelevant complications, the parameter $a_C = 1.2\; nm$ avoids the divergence character of the Coulomb potential when $x_1 = x_2$. The function $f(x_1,x_2)=exp(-(x_1^2+x_2^2)/\sigma_C)$, with $\sigma_C = 5 \; nm$, allows us to define the Coulomb interaction only in the active region of the device. These conditions mimic the solution of the 3D Poisson equation in a resonant tunneling diode with screening \cite{Oriols}. In figure~\ref{fig7}, we plot the potential $V(x_1,x_2)$ defined in equation \eref{potnosep} with $C=5$, and with the same potential barriers $V_B(x)$ discussed in \sref{sec31}. The diagonal line $x_1 = x_2$ shows the region of maximum Coulomb potential. The Coulomb potential in figure~\ref{fig7} is still symmetrical and the Left-Right symmetry is preserved.  

\begin{figure}[h]
\includegraphics[scale=0.40]{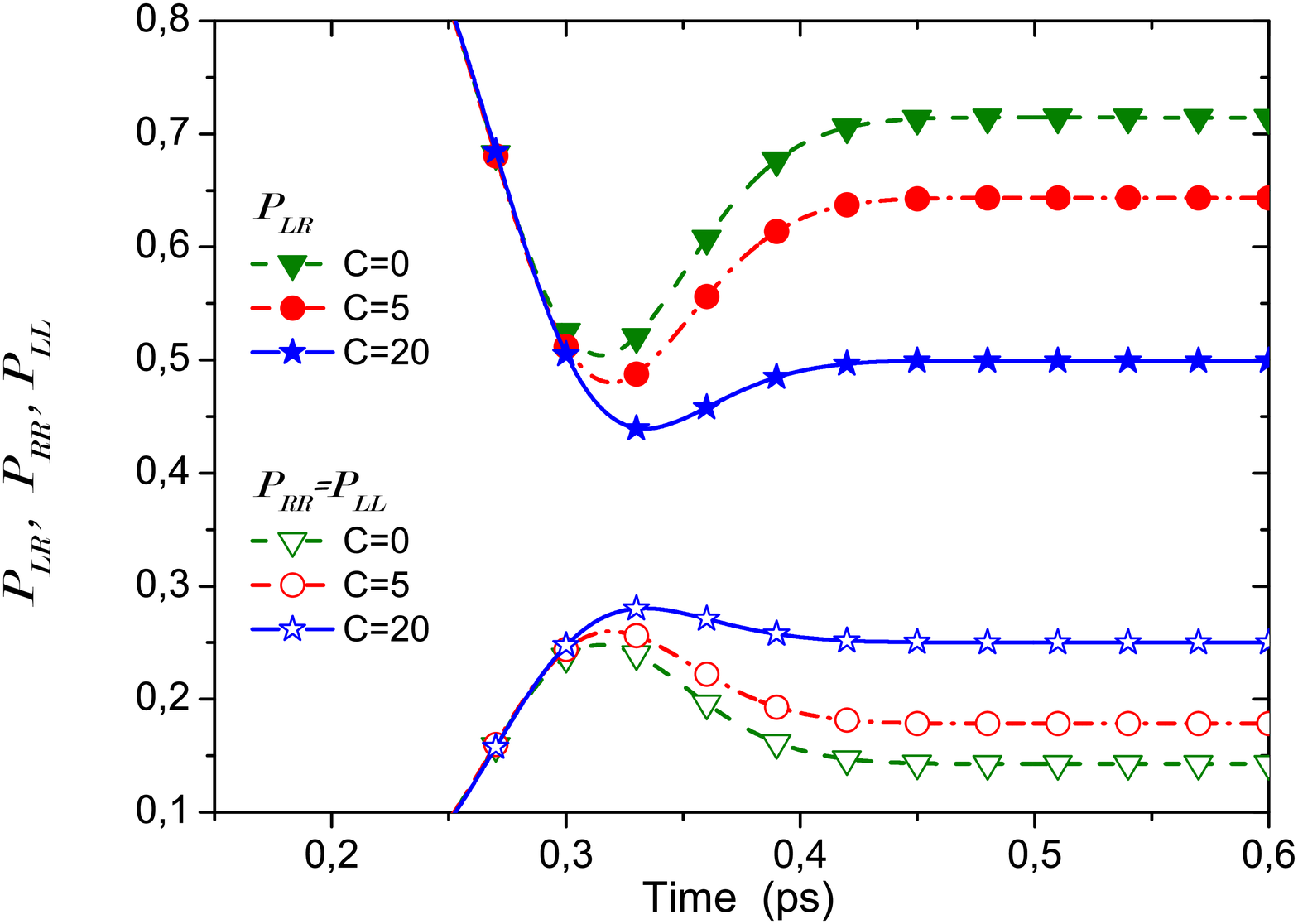}
\caption{Time evolution of $\mathcal{P}_{\mathcal{LR}}$ (upper lines) and $\mathcal{P}_{\mathcal{LL}}=\mathcal{P}_{\mathcal{RR}}$ (lower lines) from $\Phi(x_1,x_2,t)$ built by two initial wave packets located at $x_a=-175$ nm and $x_b=175$ nm with opposite momentums, equal spatial dispersions $\sigma_a=\sigma_b=35\;nm$ and equal energy $E_a = E_b = 0.069\;eV$. The values of the constant $C$ in \eref{potnosep}  are $C=0$ (green dashed line with triangles), $C=5$ (red dashed dot line with circles), and $C = 20$ (blue solid line with stars).}
\label{fig8}    
\end{figure}

\begin{figure}[h]
\includegraphics[scale=0.40]{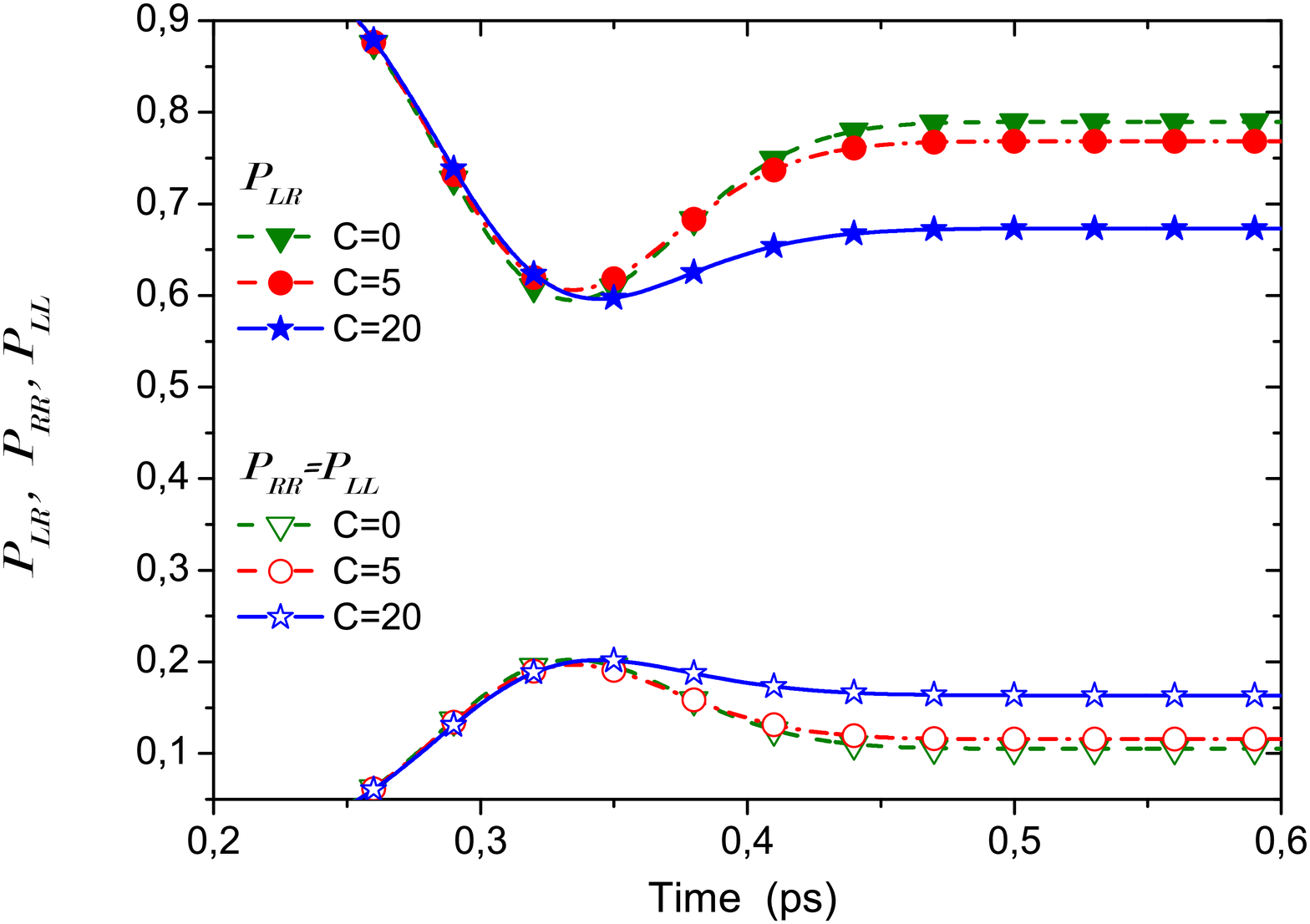}
\caption{Time evolution of $\mathcal{P}_{\mathcal{LR}}$ (upper lines) and $\mathcal{P}_{\mathcal{LL}}=\mathcal{P}_{\mathcal{RR}}$ (lower lines) from $\Phi(x_1,x_2,t)$ built by two initial wave packets located at $x_a=-175$ nm and $x_b=175$ nm with opposite momentums, equal spatial dispersions $\sigma_a=\sigma_b=35\;nm$ and equal energy $E_a = E_b = 0.06\;eV$. The values of the constant $C$ in \eref{potnosep}  are $C=0$ (green dashed line with triangles), $C=5$ (red dashed dot line with circles), and $C = 20$ (blue solid line with stars). }
\label{fig9}     
\end{figure}

In figures~\ref{fig8}~and~\ref{fig9} the two-particle probabilities for an energy  $E_a=E_b=E_R=0.069\;eV$ and  $E_a=E_b=0.06\;eV$ respectively, are plotted for different values of the constant $C$ defined in equation \eref{potnosep}. We consider exactly the same double barrier defined in \sref{sec31} with the same wave packets with $\sigma_x=35\;nm$ and $x_a=-175\; nm$ and $x_b=175\; nm$.  We conclude that the consideration of more realistic scenarios with Coulomb interaction (not directly included in the analytical computations of \ref{App0} and \ref{App1}) does not tend to recover the results $\mathcal{P}^{M}_{\mathcal{LL}}=\mathcal{P}^{S}_{\mathcal{LL}}=0$, $\mathcal{P}^{M}_{\mathcal{RR}}=\mathcal{P}^{S}_{\mathcal{RR}}=0$ and $\mathcal{P}^{M}_{\mathcal{LR}}=\mathcal{P}^{S}_{\mathcal{LR}}=1$ mentioned in equations (\ref{B_LR})-(\ref{B_RR}), but just the contrary. Again, $\mathcal{P}_{\mathcal{LR}} \neq 1$  and $\mathcal{P}_{\mathcal{LL}} = \mathcal{P}_{\mathcal{RR}} \neq 0$.


\subsection{Two-particle scenario with single barrier potential}
\label{sec4t}

\begin{figure}[h]
\includegraphics[scale=0.40]{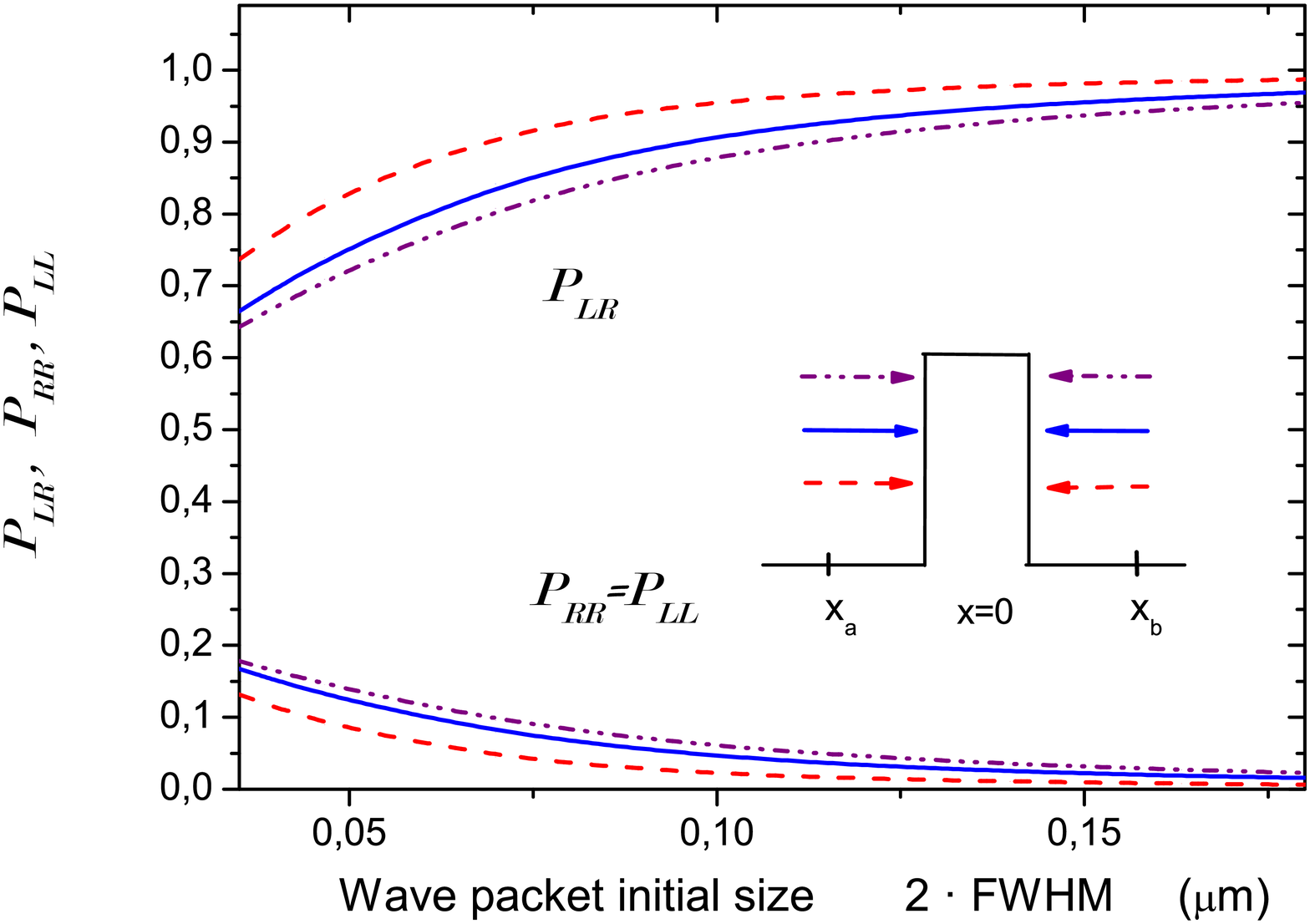}
\caption{The probabilities of $\mathcal{P}_{\mathcal{LR}}$ (upper lines)  and $\mathcal{P}_{\mathcal{LL}}=\mathcal{P}_{\mathcal{RR}}$ (lower lines) from  $\Phi(x_1,x_2,t_1)$ at time $t_1=0.8$ ps with initial wave packets located at $x_a=-175$ nm and $x_b=175$ nm with opposite momentums and with variable spatial dispersion $\sigma_x=\sigma_a=\sigma_b$. The energies are $E_a=E_b=0.035\;eV$ (red dashed line),   $E_a=E_b=E_{T=1/2}=0.045\;eV$ (blue solid line) and   $E_a=E_b=0.055\;eV$ (dash dot dot purple line). The inset shows the potential profile of a single barrier.}
\label{fig10}    
\end{figure}

One could argue that the anomalous probabilities $\mathcal{P}_{\mathcal{LL}}$ and $\mathcal{P}_{\mathcal{RR}}$ will not be accessible in a single barrier scenario because of the poorer energy dependence of the transmission and reflection coefficients. In this subsection we substitute the double barrier potential by a single barrier. This single barrier scenario is much closer to the HOM experiment mentioned in the introduction \cite{Boc,Tarucha}.  In figure~\ref{fig10} the two-particle probabilities in the case of a single barrier of width $12.4\;nm$ and height $0.04\;eV$ are plotted for three different energies as a function of the initial wave packet size. Again, only for initial wave packets with a very large spatial support (close to a Hamiltonian eigenstate) the results $\mathcal{P}^{M}_{\mathcal{LL}}=\mathcal{P}^{S}_{\mathcal{LL}}=0$, $\mathcal{P}^{M}_{\mathcal{RR}}=\mathcal{P}^{S}_{\mathcal{RR}}=0$ and $\mathcal{P}^{M}_{\mathcal{LR}}=\mathcal{P}^{S}_{\mathcal{LR}}=1$ are recovered. In particular, we plot in figure~\ref{fig10} the energy $E_a=E_b=E_{T=1/2}=0.045\;eV$ for the incident wave packets, where $E_{T=1/2}$ means that half of the wave packet is transmitted and half is reflected, in other words that the barrier works effectively as an electron beam splitter.

\begin{figure}[h]
\includegraphics[scale=0.40]{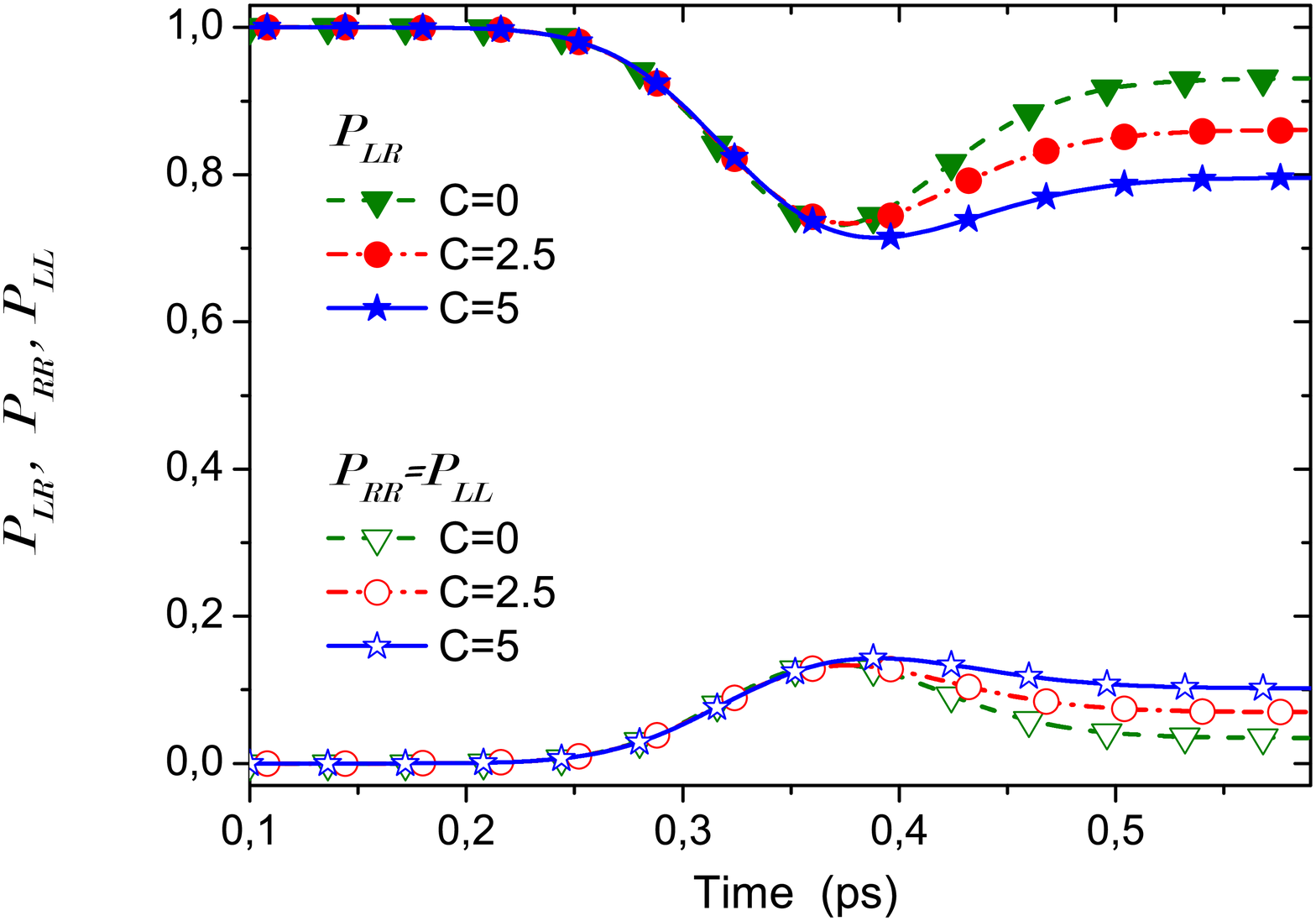}
\caption{The probabilities of $\mathcal{P}_{\mathcal{LR}}$ (upper lines)  and $\mathcal{P}_{\mathcal{LL}}=\mathcal{P}_{\mathcal{RR}}$ (lower lines) from  $\Phi(x_1,x_2,t)$ built by two initial wave packets located at $x_a=-175$ nm and $x_b=175$ nm with opposite momentums, equal spatial dispersions $\sigma_a=\sigma_b=35\;nm$ and equal energy $E_a = E_b = E_{T=1/2}=0.045\;eV$. The values of the constant $C$ in equation \eref{potnosep}  are $C=0$ (green dashed line with triangles), $C=2.5$ (red dashed dot line with circles), and $C = 5$ (blue solid line with stars).}
\label{fig11}    
\end{figure}

As shown for the double-barrier structure, also in the case of a single barrier,  the probabilities $\mathcal{P}_{\mathcal{LL}}=\mathcal{P}_{\mathcal{RR}}$ are different from zero depending on the wave packet size. The divergence from the results mentioned in equations (\ref{B_LR})-(\ref{B_RR}) is even more dramatic when considering Coulomb interaction among electrons. In figure~\ref{fig11} we report the probabilities $\mathcal{P}_{\mathcal{LR}}$, $\mathcal{P}_{\mathcal{LL}}$ and $\mathcal{P}_{\mathcal{RR}}$ as a function of time for different values of the interaction constant $C$ of equation \eref{coulpot}. We use the same values reported in \sref{sec4d}. The larger value of $C$ provides the larger discrepancies with the expected values from mono energetic states $\mathcal{P}^{M}_{\mathcal{LL}}=\mathcal{P}^{S}_{\mathcal{LL}}=0$, $\mathcal{P}^{M}_{\mathcal{RR}}=\mathcal{P}^{S}_{\mathcal{RR}}=0$ and $\mathcal{P}^{M}_{\mathcal{LR}}=\mathcal{P}^{S}_{\mathcal{LR}}=1$.


\section{Conclusions}
\label{sec5}

In this work, we consider two electrons injected simultaneously from both sides of a tunneling barrier when exchange interaction is explicitly considered. This is a typical scenario for quantum transport in electron devices and it can be also considered as a type of interference HOM experiment. We take into account explicitly quasi-particle wave packets to describe electrons. Electrons are initially associated with one-particle wave packets whose supports are located either at the left or at right of the barrier.  In the literature, such scattering experiments have been analyzed using time-independent scattering states as quasi-particles. Such states imply a probability of detecting two electrons at the same side of the barrier equal to zero \cite{Buttiker2,Buttiker3,CScho, Buttiker4}. On the contrary, in this work,  by using any type of normalizable quasi-particle wave packet as initial state, we demonstrate (analytically and numerically in many different scenarios) non-zero values for such probabilities. 

The physical origin of this non-zero probability is due to the different time-evolution suffered by the reflected and transmitted components of the wave packet. This difference between components appears in quite common scenarios (with single or double barrier potentials, with or without Coulomb interaction). For some particular resonant energies, the transmitted and reflected components are so different that they indeed become orthogonal. Then, the two-particle probabilities of these electrons with exchange interaction reproduce the probabilities predicted for distinguishable electrons. On the contrary, for initial wave packets with a large spatial support (approaching to a Hamiltonian eigenstate), the usual two-particle probabilities for indistinguishable particles reported in the literature \cite{Buttiker2,Buttiker3,CScho, Buttiker4} are exactly reproduced.  

The non-zero probabilities predicted in this work suggest a fundamental (not spurious) origin of the unexpected probabilities found in the experiments of Refs. \cite{Boc,Tarucha}. We underline that the same variability of the values of the two-particle probabilities obtained in this work can be reached from any other quantum approach, such as the scattering formalism, when it properly includes the initial state as a time-dependent wave packet. Let us notice that the difficulties calculating observables using only basis states as initial states has been also indicated in other fields, such as Rutherford scattering, \lq\lq{}because they can present misleading results when comparing to experiments.\rq\rq{}\cite{new}\\ 

Finally, we emphasize that the non-zero probabilities discussed in this work analytically (and tested numerically) have far-reaching consequences.  We notice that,  in some scenarios,  the celebrated Landauer-B\"uttiker model \cite{Buttiker2,Buttiker3,Landauer1,Landauer2} for quantum noise needs to be revisited. This model was developed within the (Landau) Fermi liquid theory \cite{Landau1,Landau2,Landau3} under the assumption of quasi-particle mono-energetic initial states. This last assumption leads to expressions (\ref{B_LR})-(\ref{B_RR}) with zero-probabilities of detecting electrons at the same side. We have explicitly shown that the consideration of quasi-particle wave packets shows non-zero probabilities in quite common scenarios. Obviously, when there are more possible scenarios for detecting the two electrons than those allowed by the Landauer-B\"uttiker model (the 4 possibilities showed in \fref{fig1}a-d including both electrons at the same place), the quantum noise is larger than what the Landauer-B\"uttiker expression predicts. In particular, we explicitly show that the unexpected non-zero probability of several particles at the same side of the barrier is experimentally accessible for a many-particle system in the limit of low phase-space density (high temperature). For high phase-space density (low temperature) the mentioned probabilities tend to zero and the fluctuation-dissipation theorem \cite{Nyquist,Johnson} is satisfied. In addition, we show that the inclusion of the Coulomb interaction between electrons (going beyond the Fermi liquid theory) does also exhibit this unexpected non-zero probabilities. An approximate algorithm to deal with the rich phenomenology of time-dependent many-particle probabilities in practical computations with exchange, tunneling and Coulomb interaction is mentioned in Refs. \cite{Oriols,alarcon1,galbareda1,galbareda2}. 

\section*{Acknowledgements}
This work has been partially supported by the \lq\lq{}Ministerio de Ciencia e Innovaci\'{o}n\rq\rq{} through the Spanish Project TEC2012-31330 and by the Grant agreement no: 604391 of the Flagship initiative  \lq\lq{}Graphene-Based Revolutions in ICT and Beyond\rq\rq{}. D.M. is supported in part by INFN and acknowledges the support of COST action (MP1006) through STSM.

\appendix
\section{Analytical two-particle probabilities from the scattering formalism with mono-energetic initial states}
\label{App0}

In this appendix we reproduce the results of the two-particle scattering probabilities for indistinguishable particles developed in Refs. \cite{Buttiker2,Buttiker3,CScho, Buttiker4} and summarized
in equations (\ref{B_LR})-(\ref{B_RR}). In the scattering formalism, input states are described by annihilation operators $\hat{a}_L$ and $\hat{a}_R$ or creation operators $\hat{a}^{\dagger}_L$ and $\hat{a}^{\dagger}_R$, being $L$ and $R$ the left and right lead. Analogously, the output states are described by $\hat{b}_i$ and $\hat{b}^{\dagger}_i$ with $i = L, R$. The connection between the $\hat{a}_i$ and the $\hat{b}_i$ is provided by the scattering matrix \cite{Buttiker2,Buttiker1}
through the relation 

\begin{equation}
{b_L \choose b_R} = \left(
\begin{array}{cc}
r & t'\\
t & r'
\end{array}
\right) {a_L \choose a_R},
\label{scatt}
\end{equation}

where the probability amplitude coefficients are such that $|r|^2=|r'|^2=R$ and $|t|^2=|t'|^2=T$ where $R$ and $T$ are respectively the reflection and transmission probabilities that satisfy  the condition $T+R=1$. Analogously the creation operators $\hat{a}^{\dagger}_i$ and the $\hat{b}^{\dagger}_i$ are related by the adjoint scattering matrix $s^{\dagger}$. The scattering matrix satisfies the relation $s^{\dagger}s=I$. For fermions the $\hat{a}_i$ operators obey the anti-commutation relations

\begin{eqnarray}
\{ \hat{a}_i, \hat{a}^{\dagger}_j\} = \delta_{ij},
\label{anticomm1}
\end{eqnarray}
and the $\hat{b}_i$ operators follows
\begin{eqnarray}
\{ \hat{b}_i, \hat{b}^{\dagger}_j\} = \delta_{ij},
\label{anticomm2}
\end{eqnarray}
with $i,j = L,R$. Equations \eref{anticomm1} and \eref{anticomm2} reflect the underling anti-symmetry of the wave function.

We focus on the physical situation depicted in figure~\ref{fig1}. An input state is constructed by one electron incident from the left and the other from the right. Both with a unique and equal (in modulus) momentum. With the help of the creation and annihilation operators we can write the input state as $|\Psi\rangle = \hat{a}^{\dagger}_{L}\hat{a}^{\dagger}_{R} |0\rangle$, where $|0\rangle$ is the vacuum state of the system. 

Using the scattering matrix (equation \eref{scatt}) and the anti-commutation relation (equations \eref{anticomm1} and \eref{anticomm2}) it is possible to obtain the probability of finding one particle on the left and the other on the right, $\mathcal{P}^{S}_\mathcal{LR}$,  as: 

\begin{eqnarray}
\mathcal{P}^{S}_\mathcal{LR}  &=& | \langle 0 | \hat{b}_{L} \hat{b}_{R} \hat{a}^{\dagger}_{L} \hat{a}^{\dagger}_{R} | 0 \rangle |^{2} = \nonumber \\
&=& | \langle 0 |  \left( r \hat{a}_{L} + t' \hat{a}_{R} \right) \left( t \hat{a}_{L} + r' \hat{a}_{R} \right) \hat{a}^{\dagger}_{L} \hat{a}^{\dagger}_{R} | 0 \rangle |^{2} = \nonumber \\
&=&   | \langle 0 |  rt \hat{a}_{L}\hat{a}_{L}\hat{a}^{\dagger}_{L}\hat{a}^{\dagger}_{R} + rr' \hat{a}_{L}\hat{a}_{R}\hat{a}^{\dagger}_{L}\hat{a}^{\dagger}_{R} + t't \hat{a}_{R}\hat{a}_{L}\hat{a}^{\dagger}_{L}\hat{a}^{\dagger}_{R} + t'r' \hat{a}_{R}\hat{a}_{R}\hat{a}^{\dagger}_{L}\hat{a}^{\dagger}_{R} | 0 \rangle |^{2} = \nonumber \\
&=& | \langle 0 | \left( rr'-t't\right) \hat{a}_{L}\hat{a}_{R}\hat{a}^{\dagger}_{L}\hat{a}^{\dagger}_{R}  | 0 \rangle |^{2} = \nonumber \\
&=&  |  \left( t't - rr' \right) |^{2} = \left( t't-rr'\right)\left( t'^*t^*-r^*r'^*\right)  =\nonumber \\
&=& T^2+R^2-t'tr^*r'^*-rr't'^*t^*  = (T+R)^2,
\label{plr-app}
\end{eqnarray}

where in the last equality of expression \eref{plr-app} we have used the property of the scattering matrix $s^{\dagger}s=ss^{\dagger}=I$.
Analogously, we can calculate two particles on the left, $\mathcal{P}^{S}_\mathcal{LL}$ , as:

\begin{eqnarray}
\mathcal{P}^{S}_\mathcal{LL}  &=& |\langle 0 |  \hat{b}_{L}  \hat{b}_{L}  \hat{a}^{\dagger}_{L} \hat{a}^{\dagger}_{R} | 0 \rangle |^{2} \nonumber 	\\
&=& | \langle 0 |  \left( r \hat{a}_{L} + t' \hat{a}_{R} \right) \left( r \hat{a}_{L} + t' \hat{a}_{R} \right) \hat{a}^{\dagger}_{L} \hat{a}^{\dagger}_{R} | 0 \rangle |^{2} = \nonumber \\
&=&   | \langle 0 |  rr \hat{a}_{L}\hat{a}_{L}\hat{a}^{\dagger}_{L}\hat{a}^{\dagger}_{R} + rt' \hat{a}_{L}\hat{a}_{R}\hat{a}^{\dagger}_{L}\hat{a}^{\dagger}_{R} + t'r \hat{a}_{R}\hat{a}_{L}\hat{a}^{\dagger}_{L}\hat{a}^{\dagger}_{R} + t't' \hat{a}_{R}\hat{a}_{R}\hat{a}^{\dagger}_{L}\hat{a}^{\dagger}_{R} | 0 \rangle |^{2} = \nonumber \\
&=& | \langle 0 | \left( rt'-t'r \right) \hat{a}_{L}\hat{a}_{R}\hat{a}^{\dagger}_{L}\hat{a}^{\dagger}_{R}  | 0 \rangle |^{2} = 0 .
\label{pll-app}
\end{eqnarray}

Finally, we can calculate the probability of detecting two particles on the right, $\mathcal{P}^{S}_\mathcal{RR}$, as:
 
 \begin{eqnarray}
\mathcal{P}^{S}_\mathcal{RR}  &=& |\langle 0 |  \hat{b}_{R}  \hat{b}_{R}  \hat{a}^{\dagger}_{L} \hat{a}^{\dagger}_{R} | 0 \rangle |^{2} \nonumber \\
&=& | \langle 0 |  \left( t \hat{a}_{L} + r' \hat{a}_{R} \right) \left( t \hat{a}_{L} + r' \hat{a}_{R} \right) \hat{a}^{\dagger}_{L} \hat{a}^{\dagger}_{R} | 0 \rangle |^{2} = \nonumber \\
&=&   | \langle 0 |  tt \hat{a}_{L}\hat{a}_{L}\hat{a}^{\dagger}_{L}\hat{a}^{\dagger}_{R} + tr' \hat{a}_{L}\hat{a}_{R}\hat{a}^{\dagger}_{L}\hat{a}^{\dagger}_{R} + r't \hat{a}_{R}\hat{a}_{L}\hat{a}^{\dagger}_{L}\hat{a}^{\dagger}_{R} + r'r' \hat{a}_{R}\hat{a}_{R}\hat{a}^{\dagger}_{L}\hat{a}^{\dagger}_{R} | 0 \rangle |^{2} = \nonumber \\
&=& | \langle 0 | \left( tr'-r't \right) \hat{a}_{L}\hat{a}_{R}\hat{a}^{\dagger}_{L}\hat{a}^{\dagger}_{R}  | 0 \rangle |^{2} = 0.
\label{prr-app}
\end{eqnarray}
Let us notice that these probabilities are developed under the assumption (implicit in the scattering formalism)  that each initial state $\hat{a}^{\dagger}_{L} |0\rangle$ or $\hat{a}^{\dagger}_{R} |0\rangle$ is a mono-energetic state. In contrast, different initial states are considered in this paper. Within the scattering formalism, a superposition of $\hat{a}^{\dagger}_{L} \hat{a}^{\dagger}_{R} |0\rangle$ with different momentums will be required to reproduce the variability of the two-particle probabilities studied in this paper.

\section{Analytical two-particle probabilities for arbitrary wave packets}
\label{App1}

A general expression for the probabilities $\mathcal{P_{LR}}$, $\mathcal{P_{LL}}$ and $\mathcal{P_{RR}}$ in equations \eref{P1}-\eref{P3} for an arbitrary normalizable wave packet is developed in this appendix. We explicitly assume the conditions (i), (ii) and (iii) mentioned in \sref{sec2}. The solution of the time dependent Schr\"odinger equation with separable potentials can be found from two decoupled single-particle Schr\"odinger equations.  After impinging with the barrier, at the time $t_1$ mentioned in the text, each initial one-particle wave function splits into (non-overlapping) transmitted ($t$) and reflected ($r$) components defined in expressions \eref{separa1} and \eref{separa2}. 

From the set of four available reflected and transmitted components, we define the set of sixteen complex integrals:
\begin{eqnarray}
I^{c,d}_{e,f}=\int_{g}^{h} dx \; \phi_e^c(x,t_1)  \; \phi_{\;f}^{*d}(x,t_1), 
\label{integral} 
\end{eqnarray}
where the upperindexes $c$ and $d$ are related to transmitted ($t$) and reflected ($r$) components, while the subindexes $e$ and $f$ to the initial position of the one-particle wave packets ($a$ left and $b$ right). The limits of the spatial integration, not explicitly indicated in $I^{c,d}_{e,f}$, are $g=-\infty,h=0$ when both components are present at the left of the barrier, while $g=0,h=\infty$ at the right. With the definitions of \eref{integral},  the transmission and reflection coefficients of the $a$-wave packet are rewritten as $T_a=I^{t,t}_{a,a}$ and $R_a=I^{r,r}_{a,a}$, respectively. Identically, we define $T_b=I^{t,t}_{b,b}$ and $R_b=I^{r,r}_{b,b}$. By construction, $I^{c,d}_{e,f}=(I^{d,c}_{f,e})^*$. 

Using the definitions in equations (\ref{separa1})-(\ref{separa2}) and (\ref{integral}), we get the property:
\begin{eqnarray}
I^{r,t}_{a,b}+I^{t,r}_{a,b}=\int_{-\infty}^{\infty} dx \; \phi_a  \; \phi_{b}^{*} = \int_{-\infty}^{\infty} dk \; g_a(k)  \; g_b^{*}(k),\;\;\;
\label{integral2} 
\end{eqnarray}
where we have defined 
\begin{eqnarray}
g_{a}(k)=\braket { \phi_{a}(x,t_0)} { \psi_k(x)}=\int_{-\infty}^{\infty}  \phi_{a}(x,t_0)  \psi_k^*(x) dx, \;\;\;\;\;\;
\end{eqnarray}
being $\psi_k(x)$ the scattering state (with $k$ its wave vector). Accordingly, the wave packet  $\phi_{a}(x,t) $ can be written by superposition as: 
\begin{eqnarray}
 \phi_{a}(x,t)=\frac {1} {\sqrt{2\pi}} \int_{-\infty}^{\infty} g_a(k)·e^{-\frac{iE_k·t} {\hbar}} \psi_k(x) dk.\;\;
\label{paquetesuperposition}
\end{eqnarray}
Identical definition for $g_b(k)$. Let us notice that the scenario depicted in figure \ref{fig1} implies that there is no overlapping between $g_{a}(k)$ and $g_{b}(k)$ because they have opposite momentums at the initial time. This no overlapping condition is true initially and it also remains valid at any later time because $\psi_k(x)$ are Hamiltonian eigenstates. Then, we get $I^{r,t}_{a,b}+I^{t,r}_{a,b}=0$. 

Using $I^{c,d}_{e,f}=(I^{d,c}_{f,e})^*$, the probability of detecting two particles at the left of the barrier in equation \eref{P4}, at $t=t_1$,  can be straightforwardly developed as:          
\begin{eqnarray}
\mathcal{P_{LL}}=\int^{0}_{-\infty} dx_1 \int^{0}_{-\infty} dx_2 \;\; |\Phi |^2=R_a T_b \mp |I^{r,t}_{a,b}|^2.
\label{P4b}
\end{eqnarray}
Identically, the probability of detecting two particles at the right of the barrier is:
\begin{eqnarray}
\mathcal{P_{RR}}=T_a R_b \mp |I^{r,t}_{a,b}|^2.
\label{P3b}
\end{eqnarray}
Finally, using also the previous identity $I^{r,t}_{a,b}=-I^{t,r}_{a,b}$, the probability of one particle at each side is:
\begin{eqnarray}
\mathcal{P_{LR}}&=&\frac {R_a R_b +T_a T_b} {2}\pm |I^{r,t}_{a,b}|^2+ \frac {R_a R_b +T_a T_b} {2}\pm |I^{r,t}_{a,b}|^2 =\nonumber \\
&=& R_a R_b +T_a T_b \pm 2 |I^{r,t}_{a,b}|^2.
\label{P1b}
\end{eqnarray}
Notice that the term $\pm |I^{r,t}_{a,b}|$ accounts for the difference between Fermions and Bosons. For these general conditions, one can check that $\mathcal{P_{LL}}+\mathcal{P_{RR}}+\mathcal{P_{LR}}=R_a R_b +T_a T_b+2T_a R_b$. Since $1=R_a+T_a$ and $1=R_b+T_b$, we finally get $\mathcal{P_{LL}}+\mathcal{P_{RR}}+\mathcal{P_{LR}}=1$, for either Fermions or Bosons.

Under the conditions (i), (ii) and (iii) mentioned in section \ref{sec2},  the expression of $I^{r,t}_{a,b}$ can be further developed. We define a new wave packet  $\Upsilon_a(x,t_1)$ as follows: $\Upsilon_a(x,t_1)=\phi_a^r(x,t_1)$ for all $x \in (-\infty,0]$ and $\Upsilon_a(x,t_1)=0$ elsewhere. This new wave packet can be written at $t_1$ as: 
\begin{eqnarray}
\Upsilon_a(x,t_1)=\frac {1} {\sqrt{2\pi}} \int_{-\infty}^{\infty} g_a(k)·e^{-\frac{iE_k·t_1} {\hbar}} {r}(k) e^{-ikx} dk,\;\;
\label{onar}
\end{eqnarray}
where ${r}(k)$ is the reflection (complex) amplitude of the scattering state $\psi_k(x)$. Notice that $\Upsilon_a(x,t_1)$ does not contain the incident plane wave $exp(ikx)$ included in $\psi_k(x)$. The reason is because, at time $t_1$, the superposition of these incident terms  $exp(ikx)$ does not contribute to the wave function at the left of the barrier. Identically, we define $\Upsilon_b(x,t_1)=\phi_b^t(x,t_1)$ for all $x \in (-\infty,0]$ and $\Upsilon_b(x,t_1)=0$ elsewhere. At $t_1$: 
\begin{eqnarray}
\Upsilon_b(x,t_1)=\frac {1} {\sqrt{2\pi}} \int_{-\infty}^{\infty} g_b(k)·e^{-\frac{iE_k·t_1} {\hbar}} {t}(k) e^{-ikx} dk,\;\;
\label{onat}
\end{eqnarray}
where ${t}(k)$ is the transmission (complex) amplitude of the scattering state $\psi_k(x)$. Because of conditions (i), (ii) and (iii), we can consider $g(k) \equiv g_a(k)=g_b(-k)$. Then, using expressions (\ref{onar}) and (\ref{onat}) we get: 
\begin{eqnarray}
I^{r,t}_{a,b}=\int_{-\infty}^{\infty} dx \Upsilon_a \Upsilon_{\;b}^{*}=\int_{-\infty}^{\infty} dk |g(k)|^2 {r}(k) {t}^*(k),
\label{integralbona} 
\end{eqnarray}
where the spatial integral in equation \eref{integralbona} extends from $-\infty$ to $\infty$ because, by construction, $\Upsilon_a(x,t_1)$ and $\Upsilon_{b}^{*}(x,t_1)$ are zero at $x \in (0,\infty)$. We have also used the property of the scattering states ${t}(k)={t}(-k)$. It is interesting to compare equation \eref{integralbona} with the well-known expression for the computation of the (one-particle) transmission coefficient: 
\begin{eqnarray}
T=T_b=T_a=I^{t,t}_{a,a}=\int_{-\infty}^{\infty} dk |g(k)|^2 |{t}(k) |^2,
\label{integralbona2} 
\end{eqnarray}
and (one-particle) reflection coefficient:
\begin{eqnarray}
R=R_b=R_a=I^{r,r}_{a,b}=\int_{-\infty}^{\infty} dk |g(k)|^2 |{r}(k)|^2.
\label{integralbona3} 
\end{eqnarray}
Notice that, under the conditions (i), (ii) and (iii), the transmission $T=T_b=T_a$ and reflection $R=R_b=R_a$ coefficients are equal for the $a$ and $b$ wave packets. We notice that $T$ and $R$ take real values, while $I^{r,t}_{a,b}$ take complex ones.   
        
From equation (\ref{integralbona}), it is a straightforward procedure to deduce the maximum allowed value for $|I^{r,t}_{a,b}|^2$. The maximum value is $|I^{r,t}_{a,b}|^2=R T$. It corresponds to an scenario where ${r}(k)$ and ${t}(k)$ are (almost) constant in the support of $g(k)$. Then, from equation \eref{integralbona}, we obtain $I^{r,t}_{a,b} \approx {r}(k^c) {t}^*(k^c)$ with $k^c$ defined as the central wave vector of the wave packet. It can be straightforwardly demonstrated that this value implies that the shapes of the $a$-reflected $\phi_r^a(x,t)$ and $b$-transmitted $\phi_t^b(x,t)$ wave packets are identical up to an arbitrary (complex) constant: 
\begin{eqnarray}
\phi_a^r(x,t_1)=\phi_b^t(x,t_1) e^{\alpha+i\beta},
\label{condition1} 
\end{eqnarray}
 being $\alpha$ and $\beta$ two real position-independent constants. For such scenarios, equations (\ref{P4b})-(\ref{P1b}) can be rewritten as $\mathcal{P}^{M}_{\mathcal{LL}}$, $\mathcal{P}^{M}_{\mathcal{RR}}$ and $\mathcal{P}^{M}_{\mathcal{RR}}$ in expressions \eref{P4M} and \eref{P1M}.  We use the upperindex $M$ in equations \eref{P4M} and \eref{P1M} to indicate that the probabilities correspond to the maximum value of $|I^{r,t}_{a,b}|^2$. We repeat that equations (\ref{P4M}) and (\ref{P1M}) exactly reproduce the results found in the literature for scattering states in Refs. \cite{Buttiker2,Buttiker3,CScho,Landauer1,Landauer2}.

However, the possibility of a minimum value $|I^{r,t}_{a,b}|^2=0$ in equation \eref{integralbona} is in general ignored in the literature. This corresponds to an scenario where ${r}(k)$ and ${t}(k)$ vary very rapidly between $1$ and $0$ on the support of $g(k)$. For example, in a sharp resonance. Then, from equation \eref{integralbona},  we get $I^{r,t}_{a,b} \approx 0$. This value means that $\phi_r^a(x,t)$ and $\phi_t^b(x,t)$ in equation \eref{integral} are orthogonal. In fact, the different schematic symbols of the wave packets in figure \ref{fig1} want to emphasize this point. When $|I^{r,t}_{a,b}|^2=0$, equations (\ref{P4b})-(\ref{P1b}) can be rewritten as $\mathcal{P}^{m}_{\mathcal{LL}}$,$\mathcal{P}^{m}_{\mathcal{RR}}$ and $\mathcal{P}^{m}_{\mathcal{LR}}$  in expressions \eref{P4m} and \eref{P1m}. We use the upperindex $m$ in expressions \eref{P4m} and \eref{P1m} to indicate that these probabilities correspond to the minimum value of $|I^{r,t}_{a,b}|^2$. The probabilities in  (\ref{P4m})-(\ref{P1m}) show no difference between indistinguishable (Fermions or Bosons) or distinguishable particles.

\section{Analytical N-particle probabilities with arbitrary wave packets}
\label{App2}

In this appendix we compute explicitly the relation \eref{proof2} between probabilities $\mathcal{P_{R^N}}$ and $\mathcal{P_{R^K}}$  defined in expression \eref{qq}. As mentioned in the text, the probability in \eref{qq} is just the determinant of the following matrix, $\mathcal{P_{R^N}} = det(M_{R^N})$

\begin{eqnarray}
M_{R^N}&= &\left(\begin{array}{cccc}  p_R(1,1) & p_R(1,2) & ... & p_R(1,N) \\ p_R(2,1) & p_R(2,2) & ... & p_R(2,N)\\ ... & ... & ... & ... \\ p_R(N,1) & p_R(N,2) & ... & p_R(N,N)  
  \end{array} \right),
 \label{matrixQQ}
\end{eqnarray} 
where the complex value $p_R(l,j)$  is a correlation function between single-particle wave packets defined in equation \eref{q}, that we rewrite here as:

\begin{eqnarray}
p_R(l,j)&=&\frac{1}{\mathcal{P_N}^{\frac{1}{N}}}\int^\infty_0\psi_l^*(x,t) \psi_j(x,t)dx \nonumber \\ &=&  \frac{1}{\mathcal{P_N}^{\frac{1}{N}}}\int^\infty_{-\infty} d^*(k)a^*_l(k) d(k)a_j(k)dk ,
\label{qapp}
\end{eqnarray}
where $d(k)$ is equal to the reflection coefficient for scattering states $r(k)$ if the initial wave packet was at the right of the barrier or equal to $t(k)$ if it was at the left side.\\

A general matrix $M$ is defined as positive-semidefinite (or sometimes nonnegative-definite) matrix if the scalar product with any vector $x$ is always non-negative. i.e.  $x^{\dag}Mx\geqslant0$. The necessary and sufficient condition to assure that the $M_{R^N}$-matrix in \eref{matrixQQ} is a positive-semidefinite matrix is that the principal minors of the $M_{R^N}$-matrix are all non-negative \cite{matrixproperties}. In fact, the first minor represents the probability of having one electron at the right side of the barrier, the second represents the probability of having two electrons and so on. Therefore, by construction these minors are all non-negative and our matrix is a positive-semidefinite matrix. Hence, their eigenvalues are all real and non-negative, since the determinant of the $M_{R^N}$-matrix is equal to the product of its eigenvalues. Next step is to rewrite the $M_{R^N}$-matrix as follows

\begin{equation}
M_{R^N}=\left(\begin{array}{cc} M_{R^{N-1}} & \alpha_{N-1} \\ \alpha_{N-1}^{\dag} & D_N \end{array} \right),
\label{Qidentity}
\end{equation}

where $\alpha_{N-1}^{\dag} = (p_R(N,1),\: p_R(N,2),\: ... ,\:p_R(N,N-1))$. We have used the property $p_R(l,j)=p_R(j,l)^*$.  Using the previous expression \eref{Qidentity} we can show the following identity:

\begin{eqnarray}
& &\left(\begin{array}{cc} I_{N-1} & 0 \\ -\alpha^{\dag}_{N-1}(M_{R^{N-1}})^{-1} & 1 \end{array} \right) 
\left(\begin{array}{cc} M_{R^{N-1}} & \alpha_{N-1} \\ \alpha^{\dag}_{N-1} & D_N \end{array} \right) \nonumber\\ & = & 
\left(\begin{array}{cc} M_{R^{N-1}} & \alpha_{N-1} \\ 0 & D_N-\alpha^{\dag}_{N-1}(M_{R^{N-1}})^{-1}\alpha_{N-1} \end{array} \right),
\label{Qidentity2}
\end{eqnarray}

where $I_{N-1}$ is just the squared identity matrix of dimension $N-1$ and $(M_{R^{N-1}})^{-1}$ the inverse matrix of $M_{R^{N-1}}$. Since the determinant of the second matrix in the left-hand side of expression \eref{Qidentity2} is just the determinant of the $M_{R^N}$-matrix (see equation \eref{Qidentity}), we obtain

\begin{eqnarray}
\!\!\!\!\!det(M_{R^N}) & = &  det(M_{R^{N-1}})(D_N-\alpha^{\dag}_{N-1}(M_{R^{N-1}})^{-1}\alpha_{N-1})\nonumber\\  & = & D_N\:det(M_{R^{N-1}})(1-\frac{\alpha^{\dag}_{N-1}(M_{R^{N-1}})^{-1}\alpha_{N-1}}{D_N}).
\label{Qdet}
\end{eqnarray}

Since the matrix $M_{R^{N}}$ is positive-semidefinite, the last term of expression \eref{Qdet} must be non-negative

\begin{equation}
1-\frac{\alpha^{\dag}_{N-1}(M_{R^{N-1}})^{-1}\alpha_{N-1}}{D_N}\geqslant0.
\label{positive}
\end{equation}

In addition, the last term on the left in equation \eref{positive} must be also positive, because the inverse of a positive-semidefinite matrix is also a positive-semidefinite matrix \cite{matrixproperties} and, by definition, we know that it accomplishes $x^{\dag}Mx\geqslant0$ for any vector $x$ (for example, $x=\alpha_{N-1}$). Therefore, equation \eref{positive} can only fluctuate between 0 and 1. In this way, we have proven that 

\begin{equation}
0\leqslant det(M_{R^N}) \leqslant D_N\: det(M_{R^{N-1}}) \leqslant 1.
\label{proof}
\end{equation}

From equation \eref{proof} it is trivial to derive equation \eref{proof2} in the text, being able to compare probabilities with different number of involved electrons.\\

The equality in \eref{proof} occurs only when the correlation of the $N$-th electron with all the others wave packets is zero, $p_R(N,j)=0$ for all $j \neq N$. In this case $\alpha_{N-1}=\bar{0}$ and the $M_{R^N}$-matrix is

\begin{equation}
M_{R^N}=\left(\begin{array}{cc} M_{R^{N-1}} & \bar{0} \\ \bar{0} & D_N \end{array} \right),
\label{Qnon-overlap}
\end{equation}
whose determinant is $det(M_{R^N}) = D_N \: det(M_{R^{N-1}})$. On the contrary,  $det(M_{R^N})=0$ when, for example, the $N$-th electron and one particular $i_0$-th electron satisfy $p_R(N,j)=p_R(i_0,j)$ for all $j \neq i_0 \neq N$ .


\section*{References}
\begin{thebibliography}{10}

\bibitem{Landau1}
Landau L D 1957 \emph{Sov. Phys. JETP} \textbf{3} 920

\bibitem{Landau2}
Landau L D 1957 \emph{Sov. Phys. JETP} \textbf{5} 101

\bibitem{Landau3}
Landau L D 1958 \emph{Sov. Phys. JETP} \textbf{8} 70

\bibitem{Buttiker2}
B\"uttiker M 1990 \emph{Phys. Rev. Lett.} \textbf{65} 2901

\bibitem{Buttiker3}
B\"uttiker M 1992 \emph{Phys. Rev. B} \textbf{46} 12485

\bibitem{HOM}
Hong C K, Ou Z Y and Mandel  L 1987  \emph{Phys. Rev. Lett.} \textbf{59} 2044 

\bibitem{CScho} 
Sch\"oenenberger C 2013 \emph{Science} \textbf{339} 1041

\bibitem{Buttiker1} 
B\"uttiker M 1999 \emph{Science} \textbf{284} 275

\bibitem{FMD}
Feldmeier H and Schnack J 2000 \emph{Rev. Mod. Phys.} \textbf{72} 655

\bibitem{FMD2}
Feldmeier H 1990 \emph{Nucl. Phys. A} \textbf{515} 147

\bibitem{FMD3}
Feldmeier H, Bieler K and Schnack J 1995 \emph{Nucl. Phys. A} \textbf{586} 493

\bibitem{FMD4}
Feldmeier H  and Schnack J 1997 \emph{Prog. Part. Nucl. Phys.} \textbf{39} 393

\bibitem{Fagas}
Fagas G, Delaney P and Greer J C 2006 \emph{Phys. Rev. B} \textbf{73} 241314

\bibitem{Oriols}
Oriols X 2007 \emph{Phys. Rev. Lett.} \textbf{98} 066803

\bibitem{Landauer1}
Martin Th and Landauer R 1992 \emph{Phys. Rev. B} \textbf{45} 1742

\bibitem{new}
Van Boxem R, Partoens B and Verbeeck J 2014 \emph{Phys. Rev. A} \textbf{89} 032715

\bibitem{Boc} 
Bocquillon E, Freulon V, Berroir J M, Degiovanni P, Pla\c{c}ais B, Cavanna A, Jin Y and Feve G 2013 \emph{Science} \textbf{339} 1054

\bibitem{Tarucha} 
Liu R C, Odom B, Yamamoto Y and Tarucha S 1988 \emph{Nature} \textbf{391} 263

\bibitem{kohn}
Heinonen O and Kohn W 1987 \emph{Phys. Rev. B} \textbf{36} 3565

\bibitem{Loudon} 
Loudon R 1998 \emph{Phys. Rev. A} \textbf{58}  4904 

\bibitem{diVentra}  
Vignale G and Di Ventra M 2009 \emph{Phys. Rev. B} \textbf{79} 014201

\bibitem{errorDFT}
Kurth S, Stefanucci G, Almbladh C O, Rubio A and Gross E K U 2005 \emph{Phys. Rev. B} \textbf{72} 035308

\bibitem {Niehaus} Yam C Y, Zheng X, Chen G H, Wang Y, Frauenheim T and Niehaus T A 2011 \emph{Phys. Rev. B} \textbf{83} 245448

\bibitem {Angel} Marques M A L, Maitra N, Nogueira F, Gross E K U and Rubio A (eds) 2012 \emph{Fundamentals of Time-Dependent Density Functional Theory, Lecture Notes in Physics}  Vol 837 (Berlin: Springer)

\bibitem{Ferry}
Oriols X and Ferry D 2013 \emph{J. Comput. Electron.} \textbf{12} 317

\bibitem{rungePRL84}
Runge E and Gross E K U 1984 \emph{Phys. Rev. Lett.} \textbf{52} 997 

\bibitem{TDDFTcurrent} 
Capelle K and Gross E K U 1997 \emph{Phys. Rev. Lett.} \textbf{78} 1872

\bibitem{varga}
Varga K 2011 \emph{Phys. Rev. B} \textbf{83} 195130

\bibitem{Bokes}
Kon\^opka M and Bokes P 2014 \emph{Phys. Rev. B} \textbf{89} 125424

\bibitem{kramer}
Kramer T, Kreisbeck C and Krueckl V 2010 \emph{Phys. Scr.} \textbf{82} 038101

\bibitem{godby}
Ramsden J D and Godby R W 2012 \emph{Phys. Rev. Lett.} \textbf{109} 036402

\bibitem{newnew}
Nussbaum I G and Kleber M 1988 \emph{J. Phys. A: Math. Gen.} \textbf{21} 2953

\bibitem{Cohen} 
Cohen-Tannoudji C, Diu B and Lalole F 1992 \emph{Quantum Mechanics} (Vol 1) (New York: Wiley)

\bibitem{Pauli} 
Pauli W 1925, Z. \emph{Physik} \textbf{31} 765

\bibitem{Buttiker4} 
Blanter M and B\"uttiker M 2000 \emph{Phys. Rep.} \textbf{336} 1

\bibitem{Landauer2} 
Landauer R and Martin T 1991 \emph{Physica B} \textbf{175} 167

\bibitem{sizes} 
Venugopal R, Goasguen S, Datta S and Lundstrom M S 2004 \emph{J. Appl. Phys.} \textbf{95} 292

\bibitem{FMD5}
Saraceno M, Kramer P and Fernandez F 1983 \emph{Nucl. Phys. A} \textbf{405} 88

\bibitem{Nyquist} 
Nyquist H 1928 \emph{Phys. Rev.} \textbf{32} 110

\bibitem{Johnson}
Johnson J 1928 \emph{Phys. Rev.} \textbf{32} 97

\bibitem{alarcon1}
Alarc\'on A, Yaro S, Cartoix\`a X and Oriols X 2013 \emph{J. Phys.: Condens. Matter.} \textbf{25} 325601

\bibitem{galbareda1}
Albareda G, Su\~{n}\'{e} J and Oriols X 2009 \emph{Phys. Rev. B} \textbf{79} 075315

\bibitem{galbareda2}
Albareda G, Lopez H, Cartoix\`{a} X, Su\~{n}\'{e} J and Oriols X 2010 \emph{Phys. Rev. B} \textbf{82} 085301

\bibitem{matrixproperties} 
Noble B 1969 \emph{Applied Linear Algebra} (Englewood Cliffs, NJ: Prentice-Hall)


\end{thebibliography}
\end{document}